\documentclass{article}
\newif\ifemail    %
\newif\ifacks     %
\newif\ifappendix %
\newif\ifedits    %
\newif\ifbanner   %
\newif\ifarxiv    %

\usepackage[preprint,nonatbib]{neurips_2023}\emailfalse\ackstrue\appendixtrue\editsfalse\bannertrue\arxivtrue %

\usepackage[utf8]{inputenc} %
\usepackage[T1]{fontenc}    %
\usepackage{url}            %
\usepackage{booktabs}       %
\usepackage{nicefrac}       %
\usepackage{microtype}      %
\usepackage{xcolor}         %

\usepackage{amsmath,amssymb,amsfonts}
\usepackage{algorithmic}
\usepackage{graphicx}
\usepackage{textcomp}
\usepackage[hidelinks]{hyperref}
\usepackage{tikz}
\usepackage[normalem]{ulem}
\usepackage{fp}
\usepackage{multirow}
\usepackage{bbding}
\usetikzlibrary{arrows.meta}
\usepackage{subcaption}
\usepackage{afterpage}
\usepackage{xspace}
\usepackage[outline]{contour}

\ifbanner\else\makeatletter\renewcommand{\@noticestring}{}\makeatother\fi
\ifappendix
  \newcommand{\refappEvaluation}[0]{\autoref{app:evaluation}}
  \newcommand{\refappResults}[0]{\autoref{app:results}}
  \newcommand{\refappMethod}[0]{\autoref{app:method}}
\else
  \newcommand{\refappEvaluation}[0]{Appendix~A}
  \newcommand{\refappResults}[0]{Appendix~B}
  \newcommand{\refappMethod}[0]{Appendix~C}
\fi

\newcommand{\PLACEHOLDER}[1]{{\color{red}{#1}}}

\newcommand{\KILL}[1]{}

\definecolor{olive}{rgb}{0.5, 0.5, 0.0}
\definecolor{maroon}{rgb}{0.69, 0.19, 0.38}
\definecolor{celestialblue}{rgb}{0.29, 0.59, 0.82}
\definecolor{darkgreen}{rgb}{0.0, 0.6, 0.0}
\definecolor{grey}{rgb}{0.5,0.5,0.5}
\definecolor{darkblue}{rgb}{0.19, 0.19, 0.62}
\definecolor{silver}{rgb}{0.7,0.7,0.7}
\definecolor{carrotorange}{rgb}{0.93, 0.57, 0.13}

\newcommand{\configA}[0]{{\small\textsc{Method~A}}\xspace}
\newcommand{\configB}[0]{{\small\textsc{Method~B}}\xspace}
\newcommand{\configC}[0]{{\small\textsc{Method~C}}\xspace}
\newcommand{\configD}[0]{{\small\textsc{Method~D}}\xspace}
\newcommand{\configE}[0]{{\small\textsc{Method~E}}\xspace}
\newcommand{\configF}[0]{{\small\textsc{Method~F}}\xspace}
\newcommand{\configG}[0]{{\small\textsc{Method~G}}\xspace}
\newcommand{\configH}[0]{{\small\textsc{Method~H}}\xspace}

\newcommand{\voltovol}[0]{volume-to-volume\xspace}
\newcommand{\voltoproj}[0]{volume-to-projection\xspace}
\newcommand{\projtovol}[0]{projection-to-volume\xspace}
\newcommand{\projtoproj}[0]{projection-to-projection\xspace}
\newcommand{\Voltovol}[0]{Volume-to-volume\xspace}

\def\clap#1{\hbox to 0pt{\hss #1\hss}}%

\ifedits
  \newcommand{\TK}[2][]{\protect\EDIT[#1]{TK}{olive}{#2}}
  
  \newcommand{\JL}[2][]{\protect\EDIT[#1]{JL}{celestialblue}{#2}}
  \newcommand{\SL}[2][]{\protect\EDIT[#1]{SL}{darkgreen}{#2}}

\else
  \newcommand{\TK}[2][]{#2}
  
  \newcommand{\JL}[2][]{#2}
  \newcommand{\SL}[2][]{#2}

\fi

\newcommand{\etal}{{et al.}}
\newcommand{\tminus}{{\textminus}}

\newcommand{\keV}{\mbox{ke\hspace*{-.08em}V}} %
\newcommand{\vparagraph}[1]{\vspace*{-1mm}\paragraph{#1}}

\newcommand{\PSNRdiff}[2]{%
\renewcommand{\PLACEHOLDER}{}%
\FPset\fa{#1}%
\FPset\fb{#2}%
\FPeval{fc}{round(fa-fb:2)}%
\FPprint{fc}%
\renewcommand{\PLACEHOLDER}[1]{{\color{red}{#1}}}%
}

\newcommand{\PSNRprint}[1]{%
\renewcommand{\PLACEHOLDER}{}%
\FPset\fa{#1}%
\FPeval{fc}{round(fa:2)}%
\FPprint{fc}%
\renewcommand{\PLACEHOLDER}[1]{{\color{red}{#1}}}%
}

\newcommand{\vv}{0mm}

\newcommand{\s}{\hphantom{0}}

\newcommand{\figPipelineOverview}{
\begin{figure*}[t]
\centering%
\includegraphics[width=\linewidth,trim={0 224 0 29},clip]{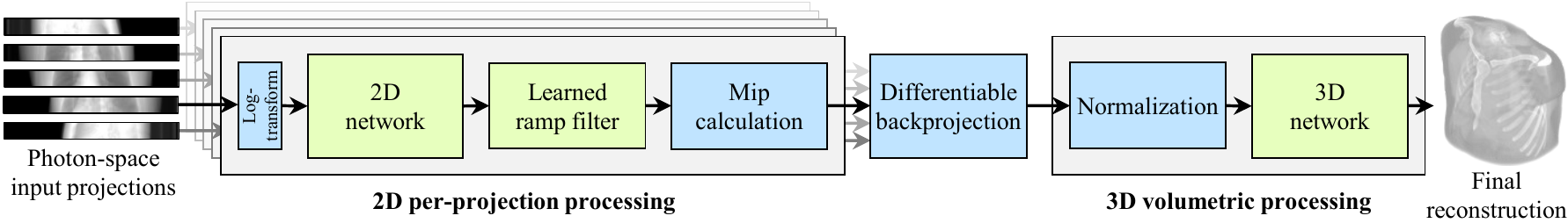}
\caption{\label{figPipelineOverview}%
Our pipeline, consisting of a combination of learned (green) and fixed-function (blue) components, reconstructs a 3D voxel volume directly from a set of raw cone-beam projections.
}
\vspace*{-1mm}
\end{figure*}
}

\newcommand{\netfigarrow}{\raisebox{9.0mm}{\begin{tikzpicture}[scale=.5]\draw[-Triangle, line width=.3mm] (0,0) -- (.75,0);\end{tikzpicture}}}
\newcommand{\netfigmacro}[1]{\fbox{\includegraphics[height=18.7mm,interpolate=false]{figures/network_in_out/#1.png}}}
\newcommand{\netfigtext}[1]{\multicolumn{1}{c}{#1}}
\newcommand{\netfigboxw}{3.4 cm} %
\newcommand{\netfigboxdep}{0.5 cm} %
\newcommand{\netfigboxh}{1.7 cm} %
\newcommand{\netfigbox}[1]{
\begin{tikzpicture}
    \setlength{\fboxrule}{0.35pt}
    \node[inner sep=0, outer sep=0, fill=black] (front)
        {\fbox{\includegraphics[width=\netfigboxw, height=\netfigboxh]{figures/network_in_out/#1_xz_000.png}}};
    \begin{scope}       
        \pgftransformxslant{1}
        \pgfset{minimum width=\netfigboxw, minimum height=\netfigboxdep, outer sep = 0}
        \pgftransformshift{\pgfpointanchor{front}{north}}
        \node[anchor=south, inner sep =0, xslant=-1, outer sep = 0, fill=black] (img) at (front.north) 
            {\fbox{\scalebox{1}[-1]{\includegraphics[width=\netfigboxw, height=\netfigboxdep]{figures/network_in_out/#1_xy_000.png}}}}; 
    \end{scope}
    \node[anchor=east, yslant=-1, inner sep=0, outer sep=0, fill=black] at (front.west)
        {\fbox{\includegraphics[width=\netfigboxdep, height=\netfigboxh]{figures/network_in_out/#1_yz_000.png}}};
\end{tikzpicture}
}
\newcommand{\figNetworkInputOutput}{
\begin{figure}[t]
\begin{subfigure}{0.254\linewidth}\centering\footnotesize%
  \includegraphics[width=\linewidth,trim={5 0 20 0},clip]{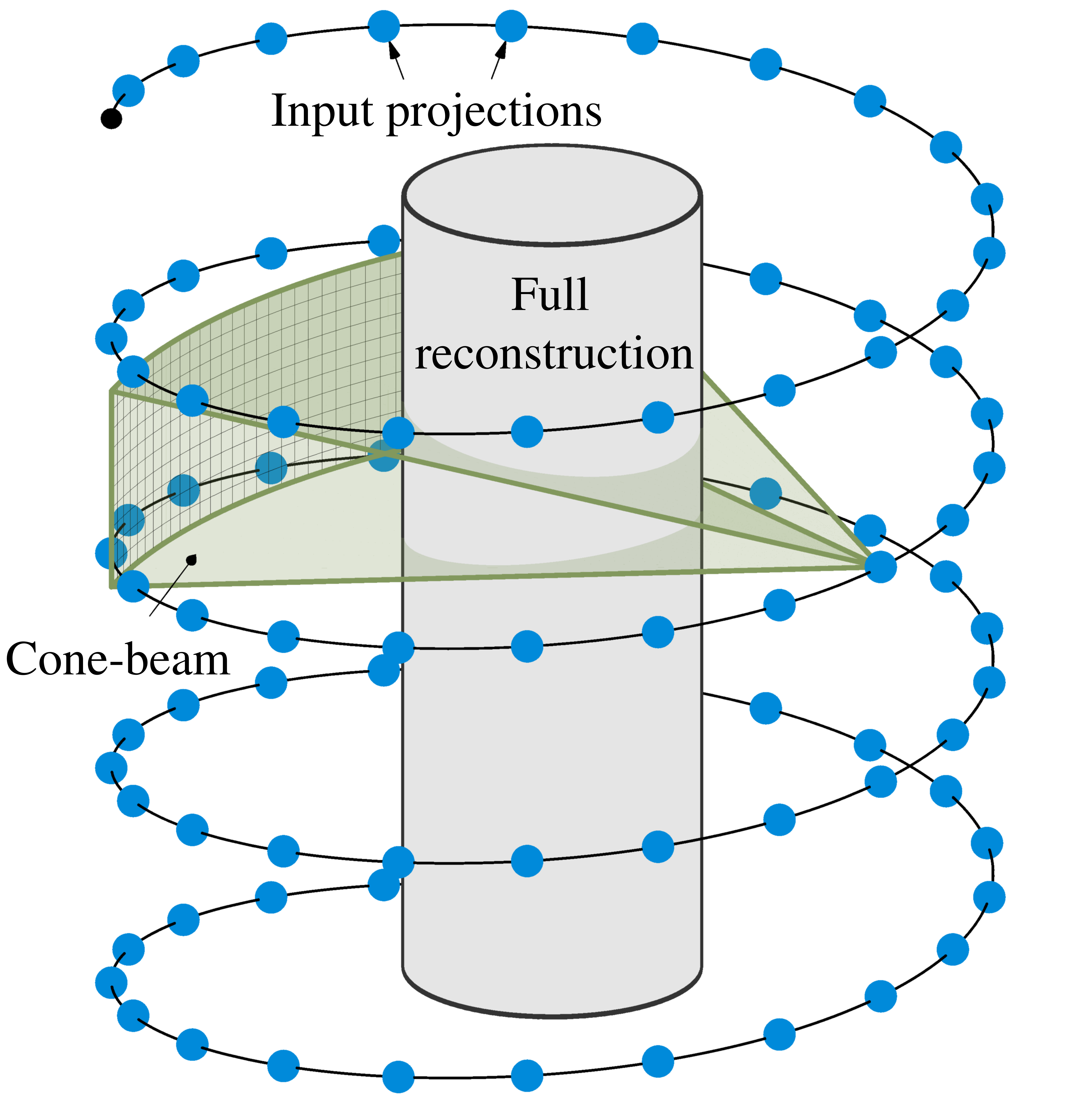}%
  \caption{Geometric setup}
\end{subfigure}\hfill%
\begin{subfigure}{0.254\linewidth}\centering\footnotesize%
  \includegraphics[width=\linewidth,trim={5 0 20 0},clip]{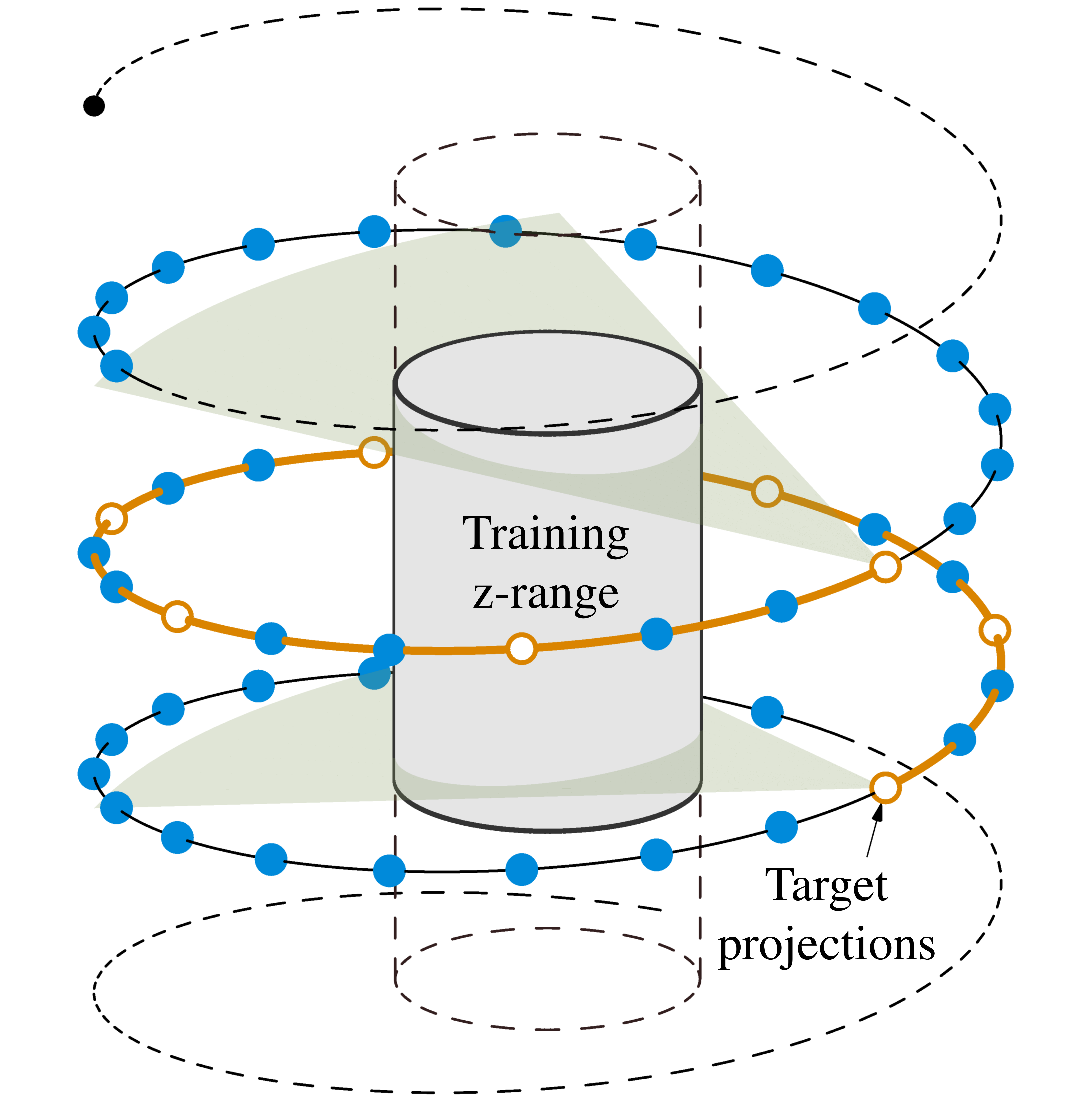}%
  \caption{Training}
\end{subfigure}\hfill%
\begin{subfigure}{0.42\linewidth}\centering\footnotesize%
  \setlength{\fboxsep}{0pt}%
  \setlength{\fboxrule}{0.6pt}%
  \resizebox{\linewidth}{!}{%
  \begin{tabular}{@{}l@{\ \ }c@{\ \ }l@{}}
    \hspace{0mm}\netfigmacro{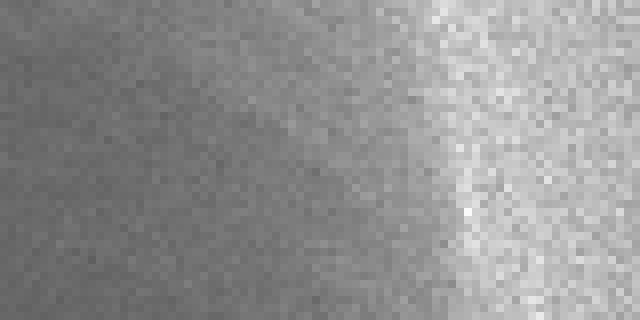} &              &  \hspace{0mm}\netfigmacro{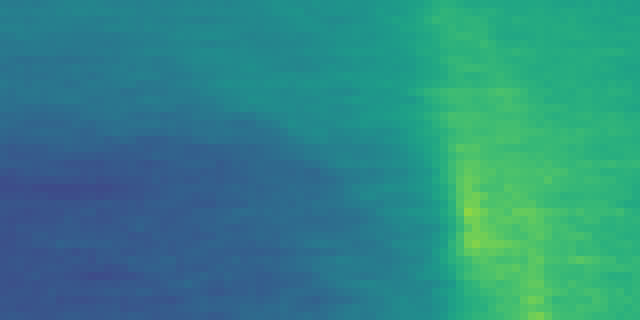} \vspace{-18.7mm}\\
    \hspace{1mm}\netfigmacro{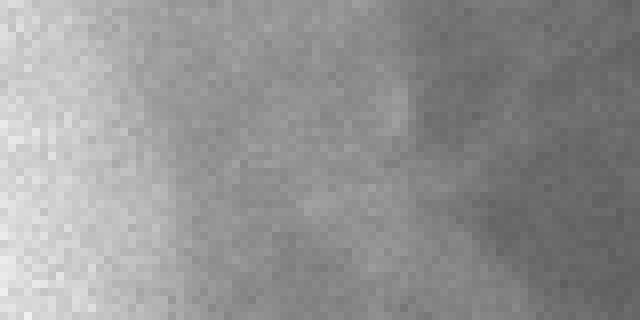} &              &  \hspace{1mm}\netfigmacro{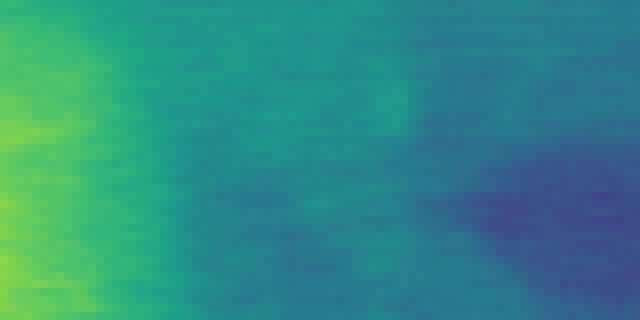} \vspace{-18.7mm}\\
    \hspace{2mm}\netfigmacro{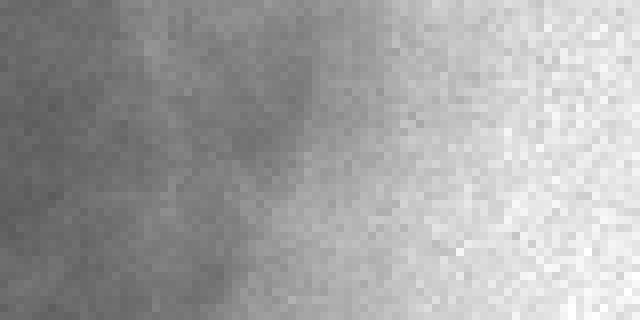} & \netfigarrow &  \hspace{2mm}\netfigmacro{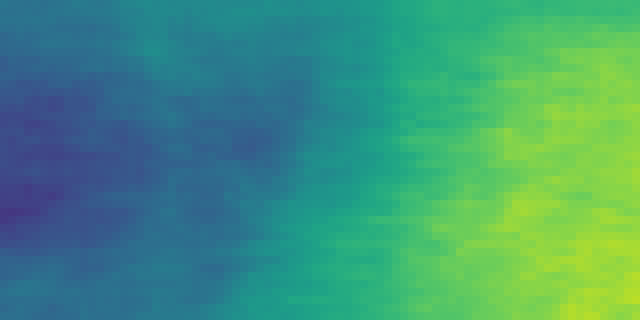} \vspace{0mm}  \\
    \netfigtext{2D network input} &\raisebox{0mm}[0mm][3mm]{}& \netfigtext{2D network output}\\
    \netfigbox{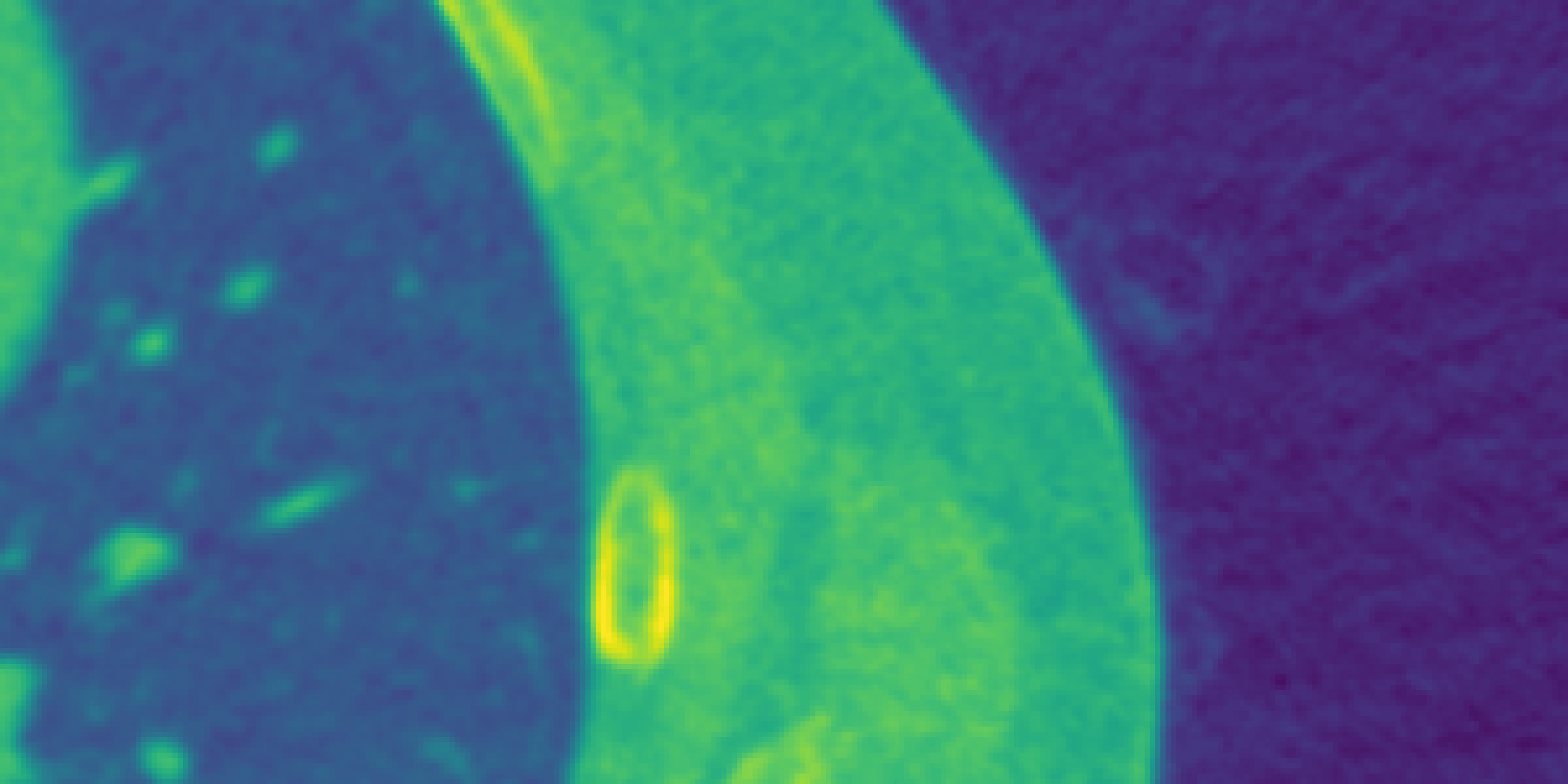} & \netfigarrow & \netfigbox{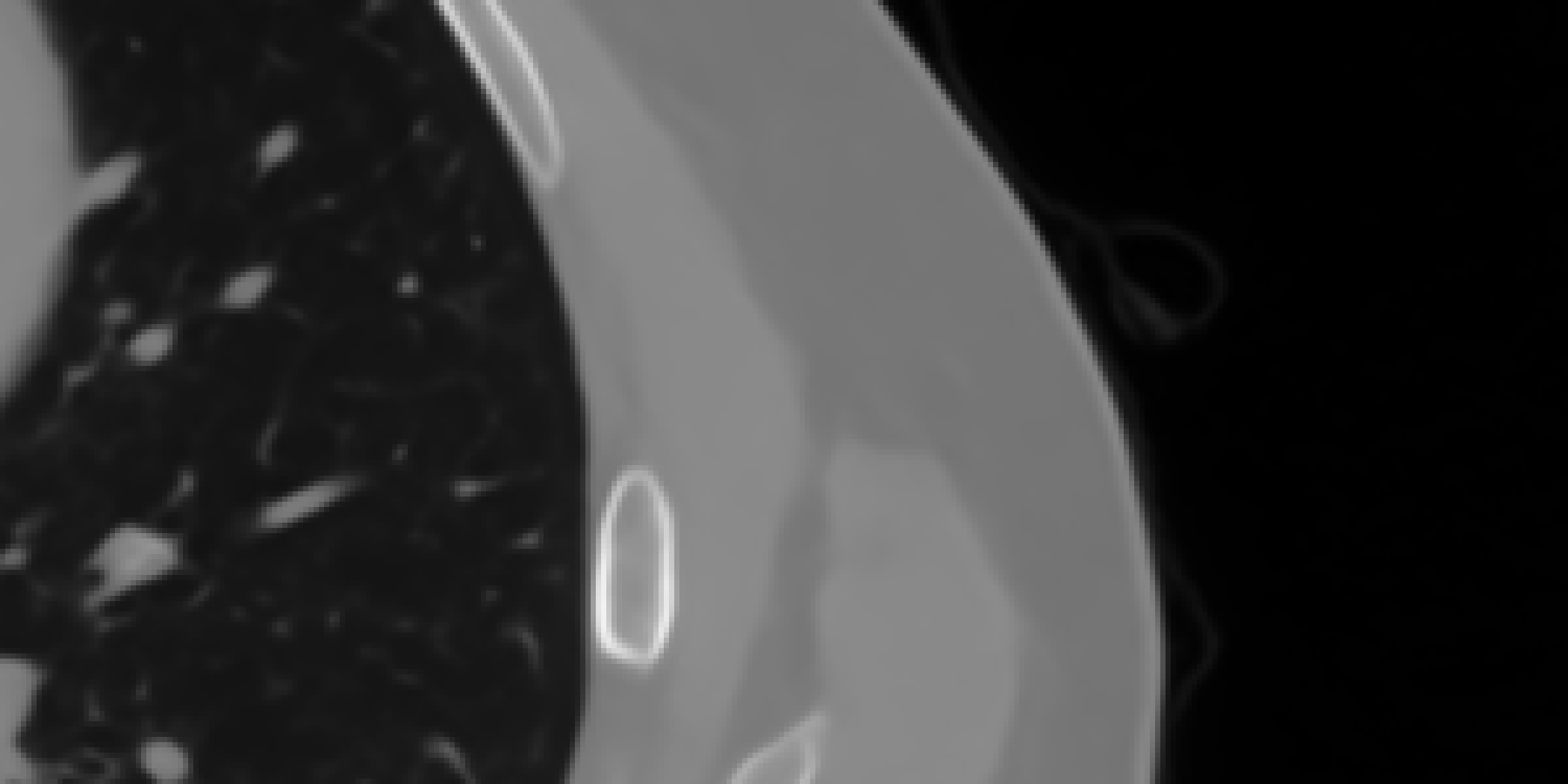} \\
    \netfigtext{3D network input} && \netfigtext{3D network output}
  \end{tabular}}%
  \caption{Inputs \& outputs of the learned networks}
\end{subfigure}%
\caption{\label{figNetworkInputOutput}%
(a) The scanner travels along a helical trajectory, capturing cone-beam projections at regular intervals with the positions of the X-ray source depicted by the blue dots. During inference, we reconstruct the full volume using all projections.
(b) For training, we randomly choose a small $z$ range (gray) corresponding to a single a rotation of the scanner, and use the projections that intersect it as either inputs (blue) or targets (orange).
(c)
The 2D network turns the log-space projections into feature maps for further processing.
The 3D network outputs the final reconstructed volume.
}
\vspace*{-1mm}
\end{figure}
}

\newcommand{\figPrefilter}{
\renewcommand{\vv}{28mm}
\begin{figure}[t]
\centering
\begin{subfigure}{1.0\linewidth}\centering\small%
  \makebox(0,0)[l]{\raisebox{14mm}{\parbox{9mm}{\centering X-ray\\source}}}%
  \includegraphics[height=\vv,trim={0 0 -0.3 0}]{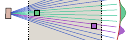}%
  \hspace{9mm}\makebox(0,0)[r]{\raisebox{22mm}{Detector}}\\
  (a) Cone-beam backprojection
\end{subfigure}\vspace*{3mm}\\
\begin{subfigure}{0.28\linewidth}\centering\small%
  \includegraphics[height=\vv,trim={0 5 0 55},clip]{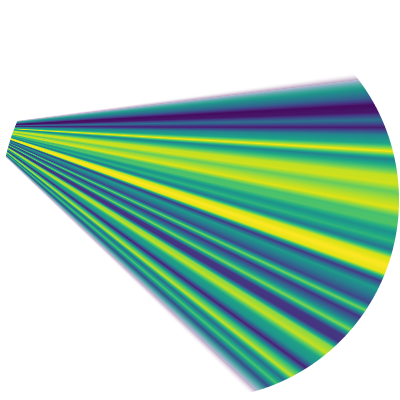}\\
  (b) Underlying signal
\end{subfigure}\hspace*{1mm}%
\begin{subfigure}{0.28\linewidth}\centering\small%
  \includegraphics[height=\vv,trim={0 5 0 55},clip]{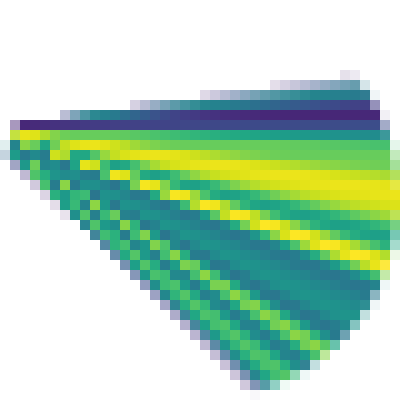}\\
  (c) Constant prefilter
\end{subfigure}\hspace*{1mm}%
\begin{subfigure}{0.28\linewidth}\centering\small%
  \includegraphics[height=\vv,trim={0 5 0 55},clip]{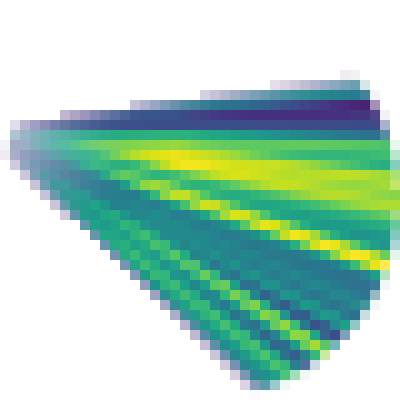}\\
  (d) Our variable prefilter
\end{subfigure}%
\caption{\label{figPrefilter}%
A variable-size prefilter is necessary for high-quality cone-beam backprojection.
(a) The green voxel near source intersects a thicker bundle of rays than the purple voxel near detector, and thus requires a wider prefilter in the projection domain to avoid aliasing.
Sampling the underlying signal (b) with a constant-size prefilter in the projection domain (c) yields aliasing near source, blurring near detector, or both as in this example.
A variable-size prefilter (d) extracts all the frequencies that the sampling grid can represent.
2D illustration, not to scale.
}
\vspace*{-1mm}
\end{figure}
}

\newcommand{\figTrainingPipeline}{
\begin{figure*}[t]
\centering%
\includegraphics[width=\linewidth,trim={0 100 0 162},clip]{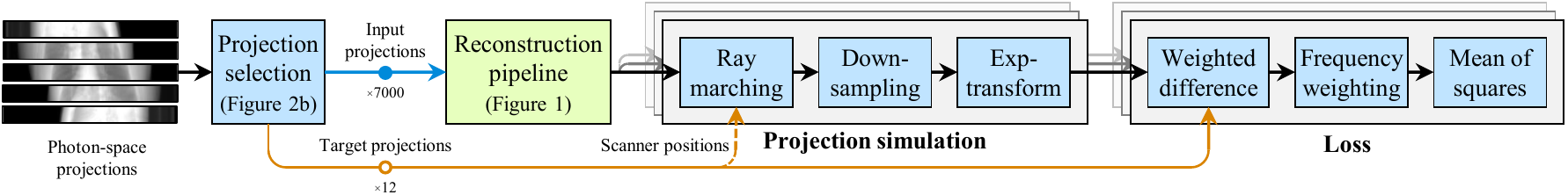}%
\caption{\label{figTrainingPipeline}%
We employ a self-supervised training setup that does not require reference data.
In each iteration, we select \TK{a handful of} target projections from a randomly chosen rotation of the scanner and feed the remaining ones as inputs to our pipeline (\autoref{figPipelineOverview}).
Based on the resulting volume, we simulate projections from the same locations as the \TK{targets}, and minimize their weighted difference.
}
\vspace*{-1mm}
\end{figure*}
}

\newcommand{\psnrFreectld}{23.23}
\newcommand{\psnrFreecthd}{33.41}
\newcommand{\psnrIRTVld}{35.49}
\newcommand{\psnrIRTVhd}{42.12}
\newcommand{\psnrFreectRedcnnTwoDSupldhd}{38.18}
\newcommand{\psnrFreectUnetTwoDSupldhd}{39.17}
\newcommand{\psnrFreectUnetThreeDSupldhd}{39.77}
\newcommand{\psnrProjnetUnetThreeDSupldhd}{39.38}
\newcommand{\psnrProjnetUnetThreeDSelfsupldld}{38.74}
\newcommand{\psnrProjnetUnetThreeDSelfsupldhd}{41.11}
\newcommand{\psnrProjnetUnetThreeDSelfsuphdhd}{44.21}
\newcommand{\psnrProjnetUnetThreeDSelfsupldldLogLoss}{35.42}
\newcommand{\rmseFreectld}{0.1379}
\newcommand{\rmseFreecthd}{0.0427}
\newcommand{\rmseIRTVld}{0.0336}
\newcommand{\rmseIRTVhd}{0.0157}
\newcommand{\rmseFreectRedcnnTwoDSupldhd}{0.0247}
\newcommand{\rmseFreectUnetTwoDSupldhd}{0.0220}
\newcommand{\rmseFreectUnetThreeDSupldhd}{0.0206}
\newcommand{\rmseProjnetUnetThreeDSelfsupldhd}{0.0176}
\newcommand{\rmseProjnetUnetThreeDSelfsuphdhd}{0.0123}
\newcommand{\runtimeFreectld}{90s}

\newcommand{\runtimeIRTVld}{1h}

\newcommand{\runtimeFreectRedcnnTwoDSupldhd}{106s}
\newcommand{\runtimeFreectUnetTwoDSupldhd}{92s}
\newcommand{\runtimeFreectUnetThreeDSupldhd}{93s}
\newcommand{\runtimeProjnetUnetThreeDSelfsupldhd}{27s}
\newcommand{\ch}[1]{\vphantom{$\sqrt{A}_j$}\textbf{#1}}
\newcommand{\sliced}{$^\ast$}
\newcommand{\needsref}{$^\dag$}
\newcommand{\rstrut}{\vphantom{\needsref{}}} %
\newcommand{\tabXcatMainResultsTable}{
\begin{table}[t]
\caption{\label{tabXcatMainResultsTable}%
Quantitative reconstruction quality of different methods using synthetic data.
}
\vspace{1mm}
\centering\footnotesize%
\begin{tabular}{|l@{\hspace{2mm}}l@{\hspace{-3mm}}c@{\hspace{1.5mm}}c@{\hspace{2mm}}c|c@{\hspace{2mm}}c@{\hspace{4mm}}|c@{\hspace{2mm}}c@{\hspace{4mm}}|c|}
\hline
\multicolumn{5}{|c|}{\ch{Method}}
  & \multicolumn{2}{c|}{\ch{Low-dose inputs}}
  & \multicolumn{2}{c|}{\ch{Full-dose inputs}}
  & \ch{Runtime} \\[0.5mm]
& Pipeline & Self-sup. & Input & Target
  & PSNR~(dB) & RMSE
  & PSNR~(dB) & RMSE
  & \\
\hline
\textsc{a}\rstrut{} & wFBP~\cite{Stierstorfer04} & & Proj. & --
  & \psnrFreectld & \rmseFreectld
  & \psnrFreecthd & \rmseFreecthd
  & \s\runtimeFreectld \\
\textsc{b}\rstrut{} & IR-TV~\cite{Sidky08} & & Proj. & --
  & \psnrIRTVld & \rmseIRTVld
  & \psnrIRTVhd & \rmseIRTVhd
  & $\sim$\runtimeIRTVld \\
\hline
\textsc{c}\rstrut{} & RED-CNN~\cite{Chen17}\sliced{} & & Vol.\needsref{} & Vol.\needsref{}
  & \psnrFreectRedcnnTwoDSupldhd & \rmseFreectRedcnnTwoDSupldhd
  & -- & --
  & \runtimeFreectRedcnnTwoDSupldhd \\
\textsc{d}\rstrut{} & Our 2D U-Net\sliced{} & & Vol.\needsref{} & Vol.\needsref{}
  & \psnrFreectUnetTwoDSupldhd & \rmseFreectUnetTwoDSupldhd
  & -- & --
  & \s\runtimeFreectUnetTwoDSupldhd \\
\textsc{e}\rstrut{} & Our 3D U-Net & & Vol.\needsref{} & Vol.\needsref{}
  & \psnrFreectUnetThreeDSupldhd & \rmseFreectUnetThreeDSupldhd
  & -- & --
  & \s\runtimeFreectUnetThreeDSupldhd \\
\textsc{f}\rstrut{} & Our pipeline & & Proj. & Vol.\needsref{}
  & \psnrProjnetUnetThreeDSupldhd & 0.0215
  & -- & --
  & \s\runtimeProjnetUnetThreeDSelfsupldhd \\
\hline
\textsc{g}\rstrut{} & \ch{Our pipeline} & \hspace{2mm}\checkmark & Proj. & Proj.
  & \ch{\psnrProjnetUnetThreeDSelfsupldhd} & \ch{\rmseProjnetUnetThreeDSelfsupldhd}
  & \ch{\psnrProjnetUnetThreeDSelfsuphdhd} & \ch{\rmseProjnetUnetThreeDSelfsuphdhd}
  & \s\ch{\runtimeProjnetUnetThreeDSelfsupldhd} \\
\textsc{h}\rstrut{} & Our 3D U-Net & \hspace{2mm}\checkmark & Vol.\needsref{} & Proj.
  & 39.59 & 0.0210
  & -- & --
  & \s\runtimeFreectUnetThreeDSupldhd \\
\hline
\multicolumn{10}{c}{}\\[-3mm]
\multicolumn{6}{@{}l}{\scalebox{0.9}{\sliced{} Processes each $z$-slice of the volume independently.}} &
\multicolumn{4}{r@{}}{\scalebox{0.9}{\needsref{} Relies on wFBP to obtain the inputs and/or targets.}}\\
\end{tabular}%
\end{table}
}

\newcommand{\figLdctMainResultsFigure}{
\begin{figure*}[t]
\centering\footnotesize%
\makebox[46mm][c]{Reference}%
\makebox[18.7mm][c]{Zoomed}%
\makebox[18.7mm][c]{wFBP (\textsc{a})}%
\makebox[18.7mm][c]{IR-TV (\textsc{b})}%
\makebox[18.7mm][c]{3D\,U-Net (\textsc{e})}%
\makebox[18.7mm][c]{\textbf{Ours} (\textsc{g})}\\
\includegraphics[width=\linewidth,interpolate=false,trim={28 0 2 22.5},clip]{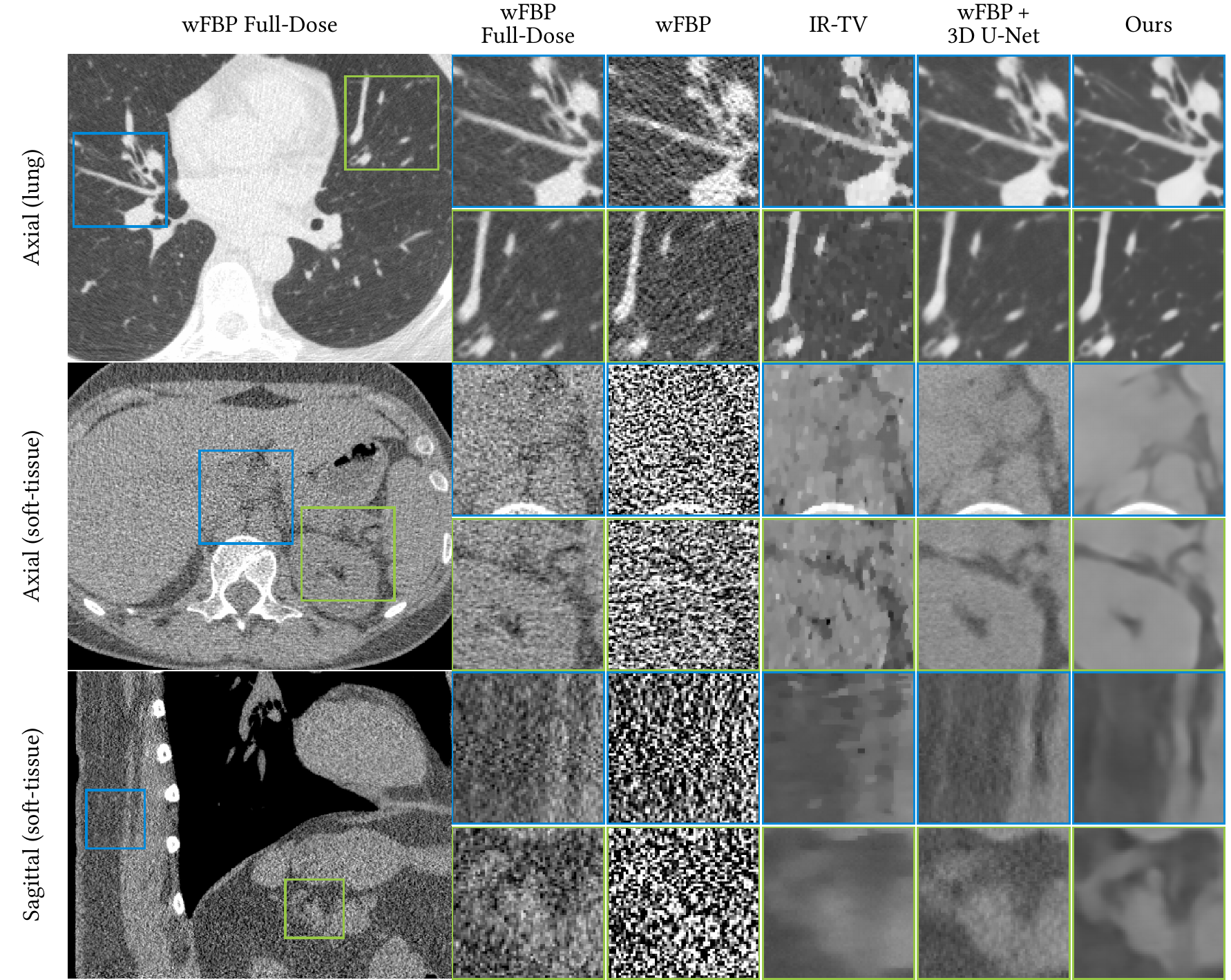}%
\makebox(0,0)[l]{\hspace{-\linewidth}\hspace{1mm}\raisebox{106mm}{\rotatebox{90}{%
  \color{white}\contourlength{0.3mm}\contour{black}{Sagittal (soft tissue)\hspace{14mm}Axial (soft tissue)\hspace{18mm}Axial (lung)}%
}}}%
\caption{\label{figLdctMainResultsFigure}%
Low-dose reconstructions of real-world data with different methods.
For reference, full-dose wFBP (two leftmost columns) is shown in lieu of noise-free ground truth that is not available.
Soft tissue display intensity range (window) is set to \mbox{[\tminus300, 300]} Hounsfield units (HU) and lung window to \mbox{[\tminus1350, 150] HU}.
Full images and neighboring slices are available in the supplement.
}
\vspace*{-1mm}
\end{figure*}
}

\newcommand{\figHdResults}{
\begin{figure*}[t]
\centering\footnotesize%
\makebox[70.2mm][c]{Full-dose reconstruction using wFBP (\textsc{a})}%
\makebox[23.1mm][c]{Zoomed}%
\makebox[23.1mm][c]{IR-TV (\textsc{b})}%
\makebox[23.1mm][c]{\textbf{Ours} (\textsc{g})}\\
\includegraphics[width=\linewidth,interpolate=false,trim={28 0 2 22.5},clip]{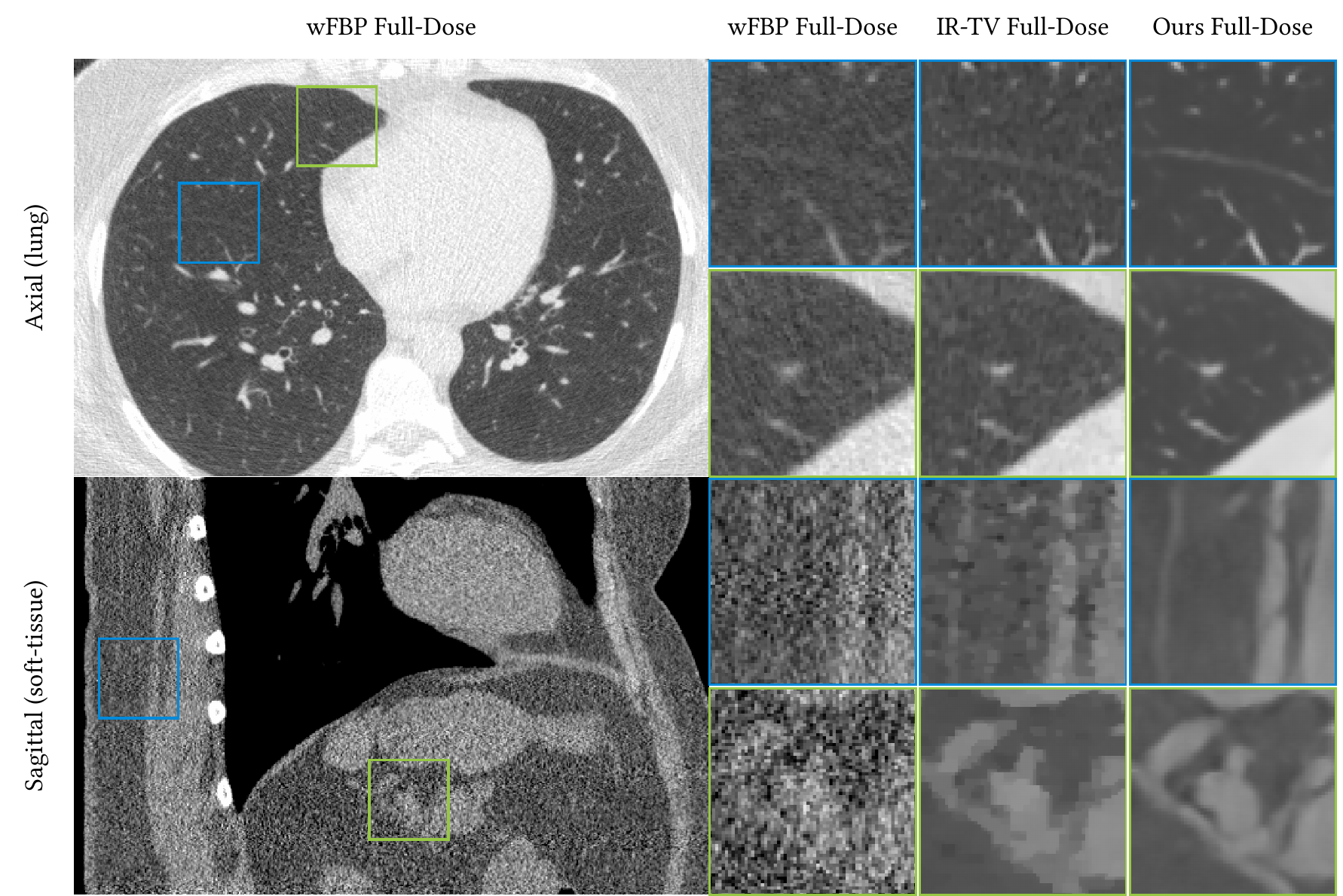}%
\makebox(0,0)[l]{\hspace{-\linewidth}\hspace{1mm}\raisebox{86mm}{\rotatebox{90}{%
  \color{white}\contourlength{0.3mm}\contour{black}{Sagittal (soft tissue)\hspace{27mm}Axial (lung)}%
}}}%
\caption{\label{figHdResults}%
Full-dose reconstructions of real-world data.
Soft tissue window is set to \mbox{[\tminus300, 300] HU}, and lung window to \mbox{[\tminus1350, 150]} HU.
Our self-supervised loss enables us to train our pipeline with full-dose input data, which is not possible with traditional supervised methods.
Compared to FBP and total variation regularized iterative reconstruction, our results contain less noise and do not suffer from IR-TV's blockiness.
Full images and neighboring slices are available in the supplement.
}
\vspace*{-1mm}
\end{figure*}
}

\newcommand{\figXcatMainResultsFigure}{
\begin{figure*}[t]
\centering\footnotesize%
\makebox[46mm][c]{Ground truth}%
\makebox[18.7mm][c]{Zoomed}%
\makebox[18.7mm][c]{wFBP (\textsc{a})}%
\makebox[18.7mm][c]{IR-TV (\textsc{b})}%
\makebox[18.7mm][c]{3D\,U-Net (\textsc{e})}%
\makebox[18.7mm][c]{\textbf{Ours} (\textsc{g})}\\
\includegraphics[width=\linewidth,interpolate=false,trim={28 0 2 22.5},clip]{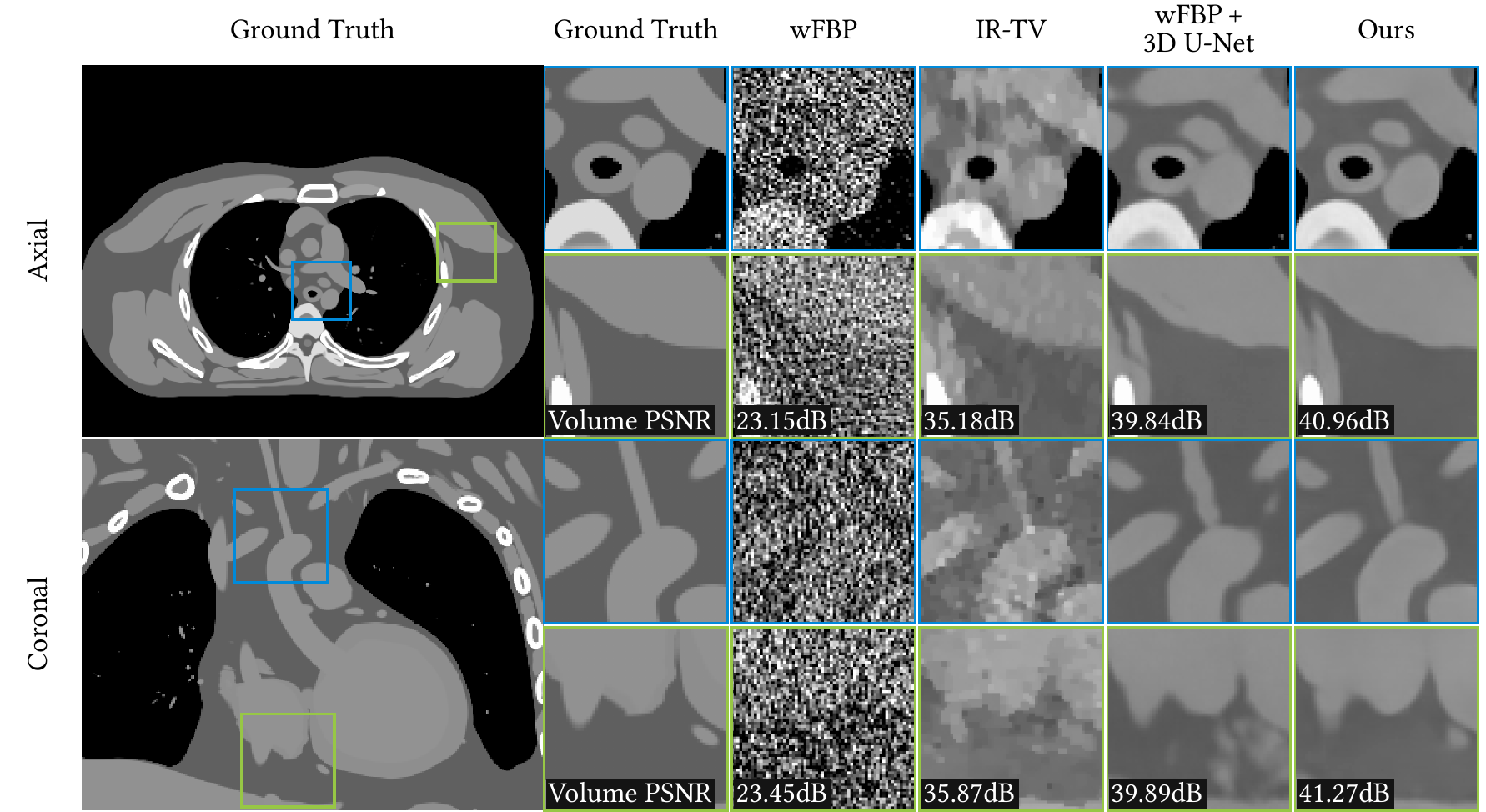}%
\makebox(0,0)[l]{\hspace{-\linewidth}\hspace{1mm}\raisebox{72mm}{\rotatebox{90}{%
  \color{white}\contourlength{0.3mm}\contour{black}{Coronal\hspace{29mm}Axial}%
}}}%
\caption{\label{figXcatMainResultsFigure}%
Low-dose reconstructions of synthetic data with different methods.
Display window is set to \mbox{[\tminus400, 400] HU}.
PSNR values refer to the individual volumes shown, not the entire validation set.
Full images and neighboring slices are available in the interactive HTML viewer.
}
\vspace*{-1mm}
\end{figure*}
}

\newcommand{\figXcatTwoDvsThreeD}{
\begin{figure}[t]
\centering\footnotesize%
\makebox[44.8mm][c]{RED-CNN (\textsc{c})}%
\makebox[22.4mm][c]{Zoomed}%
\makebox[22.4mm][c]{2D\,U-Net (\textsc{d})}%
\makebox[22.4mm][c]{3D\,U-Net (\textsc{e})}\\
\includegraphics[width=0.8\linewidth,trim={28 0 2 22},clip,interpolate=false]{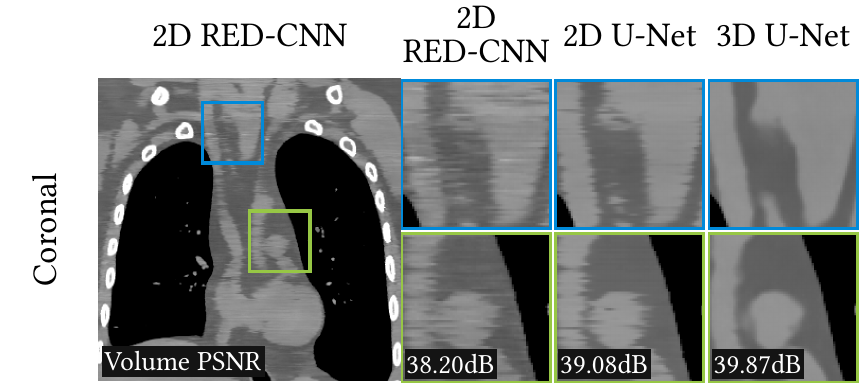}%
\caption{\label{figXcatTwoDvsThreeD}%
Coronal slices of low-dose reconstructions of synthetic data using supervised 2D and 3D denoisers.
As the 2D methods process each axial slice independently, they suffer from inconsistencies in other planes.
A 3D denoiser is consistent along all axes.
}
\vspace*{-1mm}
\end{figure}
}

\newcommand{\targetexfigmacro}[1]{\includegraphics[width=0.35\linewidth,interpolate=false]{figures/target_exclusion/#1.png}}
\newcommand{\figTargetExclusion}{
\begin{figure}[t]
\centering\footnotesize%
\begin{tabular}{@{}c@{\hspace{10mm}}c@{}}
\targetexfigmacro{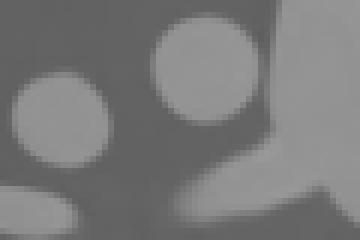} &
\targetexfigmacro{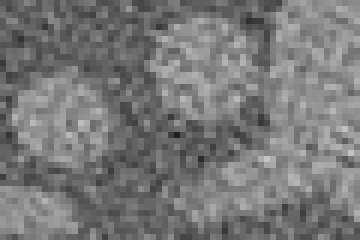} \\
\parbox[t]{41mm}{\centering (a) Target projections correctly excluded from the input set} &
\parbox[t]{41mm}{\centering (b) Target projections incorrectly included in the input set} \\
\end{tabular}
\caption{\label{figTargetExclusion}%
Example crops of our method trained with target projections excluded (a) and included (b) in the input set, using synthetic full-dose data.
If the targets are not excluded, the network input becomes correlated with the targets and the networks learn to pass the noise through instead of removing it.
Display window is set to \mbox{[\tminus400, 400] HU}.
}
\vspace*{-1mm}
\end{figure}
}

\newcommand{\dstrut}{\vphantom{$\sqrt{A}$}}
\newcommand{\dds}{\hspace*{.1em}--\hspace*{.1em}}
\newcommand{\ddt}{\hspace*{.1em}--}
\newcommand{\mm}{\,\scalebox{0.95}{mm}}
\newcommand{\tabDatasetTable}{
\begin{table}[t]
\caption{\label{tabDatasetTable}%
Dataset details.
}
\vspace{1mm}
\small
\centering%
\begin{tabular}{|l@{\hspace{4mm}}|@{\hspace{4mm}}c@{\hspace{4mm}}|@{\hspace{4mm}}c@{\hspace{4mm}}|}
\hline
\ch{Dataset details}                & \ch{Real-world}   & \ch{Synthetic}        \\
\hline
\dstrut{}Source                     & LDCT~\cite{ldct}  & XCAT~\cite{Segars08}  \\
Training scans                      & 41                & 13                    \\
Validation scans                    & 6                 & 4                     \\
Projections per scan                & 11,000\ddt15,000  & 9,000\ddt12,000       \\
Projection resolution               & 736$\times$64     & 736$\times$64         \\
Spiral pitch                        & 0.9               & 0.9                   \\
\hline
\dstrut{}Scan range                 & 240\dds380\mm     & 210\dds300\mm         \\
$xy$ voxel grid size                & 1024$\times$1024  & 576$\times$576        \\
$z$ voxel grid size (training)      & 160               & 128                   \\
$z$ voxel grid size (inference)     & 402\dds644        & 272\dds374            \\
Reconstruction voxel spacing        & 0.586\mm          & 0.784\mm              \\
Reconstruction cylinder diameter    & 600\mm            & 452\mm                \\
\hline
\end{tabular}
\end{table}
}

\newcommand{\layer}[1]{\small\textsc{#1}}
\newcommand{\convsize}{$3\times3\hspace{.1em}(\!\!{}\times3)$}
\newcommand{\maxpoolsize}{$2\times2\hspace{.1em}(\!\!{}\times2)$}
\newcommand{\ninsize}{$1\times1\hspace{.1em}(\!\!{}\times1)$}
\newcommand{\cc}[1]{\multicolumn{2}{l|}{#1}}
\newcommand{\bracketbase}{\scalebox{0.8}[1.0]{\raisebox{4ex}[0pt][0pt]{\rlap{\,$\left.\begin{matrix}\vspace*{7ex}\end{matrix}\right)$}}}}
\newcommand{\bbracket}{\scalebox{1.0}[0.45]{\bracketbase}}
\newcommand{\bbbracket}{\scalebox{1.0}[0.7]{\bracketbase}}
\newcommand{\bbbbracket}{\scalebox{1.0}[0.94]{\bracketbase}}
\newcommand{\bbbbbracket}{\rlap{\hspace*{.25em}\scalebox{1.0}[1.2]{\bracketbase}}}
\newcommand{\tabNetworkTable}{
\begin{table}[t]
\caption{\label{tabNetworkTable}%
Architecture of our neural networks.
Columns $N_\mathrm{2D}$ and $N_\mathrm{3D}$ show the output channel count of each layer in our projection and volume networks, respectively.
Joining arcs in the latter column indicate gradient checkpointing blocks in the volume network, whereas the projection network can be thought of as a single large checkpointing block.}
\centering
\vspace{2mm}
\begin{tabular}{|l|c|c@{\ \ \ }|l@{\ \ }l|}
\hline
\textsc{Name\vphantom{$A^A$}} & $N_\mathrm{2D}$ & $N_\mathrm{3D}$ & \cc{\textsc{Function}} \\
\hline
\raisebox{0mm}[3mm]{}%
\layer{input}       & $1$    & $1$   &             &                            \\
\layer{enc\_conv0}  & $48$   & $24$  & Convolution & \convsize                  \\
\layer{enc\_conv1}  & $48$   & $24$  & Convolution & \convsize                  \\
\layer{pool1}       & $48$   & $24$\bbbracket  & Max-pool    & \maxpoolsize               \\
\layer{enc\_conv2}  & $48$   & $24$  & Convolution & \convsize                  \\
\layer{pool2}       & $48$   & $24$\bbracket  & Max-pool    & \maxpoolsize               \\
\layer{enc\_conv3}  & $48$   & $24$  & Convolution & \convsize                  \\
\layer{pool3}       & $48$   & $24$  & Max-pool    & \maxpoolsize               \\
\layer{enc\_conv4}  & $48$   & $24$  & Convolution & \convsize                  \\
\layer{pool4}       & $48$   & $24$  & Max-pool    & \maxpoolsize               \\
\layer{enc\_conv5}  & $48$   & $24$  & Convolution & \convsize                  \\
\layer{pool5}       & $48$   & $24$  & Max-pool    & \maxpoolsize               \\
\layer{enc\_conv6}  & $48$   & $24$  & Convolution & \convsize                  \\
\layer{upsample5}   & $48$   & $24$  & Upsample    & \maxpoolsize               \\
\layer{concat5}     & $96$   & $48$  & \cc{Concatenate output of \layer{pool4}} \\
\layer{dec\_conv5a} & $96$   & $48$  & Convolution & \convsize                  \\
\layer{dec\_conv5b} & $96$   & $48$  & Convolution & \convsize                  \\
\layer{upsample4}   & $96$   & $48$  & Upsample    & \maxpoolsize               \\
\layer{concat4}     & $144$  & $72$  & \cc{Concatenate output of \layer{pool3}} \\
\layer{dec\_conv4a} & $96$   & $48$  & Convolution & \convsize                  \\
\layer{dec\_conv4b} & $96$   & $48$  & Convolution & \convsize                  \\
\layer{upsample3}   & $96$   & $48$  & Upsample    & \maxpoolsize               \\
\layer{concat3}     & $144$  & $72$  & \cc{Concatenate output of \layer{pool2}} \\
\layer{dec\_conv3a} & $96$   & $48$  & Convolution & \convsize                  \\
\layer{dec\_conv3b} & $96$   & $48$\bbbbracket  & Convolution & \convsize                  \\
\layer{upsample2}   & $96$   & $48$  & Upsample    & \maxpoolsize               \\
\layer{concat2}     & $144$  & $72$  & \cc{Concatenate output of \layer{pool1}} \\
\layer{dec\_conv2a} & $96$   & $48$  & Convolution & \convsize                  \\
\layer{dec\_conv2b} & $96$   & $48$\bbbbracket  & Convolution & \convsize                  \\
\layer{upsample1}   & $96$   & $48$  & Upsample    & \maxpoolsize               \\
\layer{concat1}     & $97$   & $49$  & \cc{Concatenate \layer{input}}           \\
\layer{dec\_conv1a} & $64$   & $32$  & Convolution & \convsize                  \\
\layer{dec\_conv1b} & $32$   & $16$  & Convolution & \convsize                  \\
\layer{dec\_conv1c} & $1$    & $1$\bbbbbracket   & Convolution & \ninsize, linear act.      \\
\hline
\end{tabular}
\end{table}
}

\newcommand{\figXcatLdToLd}{
\begin{figure*}[t]
\centering\footnotesize%
\makebox[53.4mm][c]{Ground truth}%
\makebox[21.6mm][c]{Zoomed}%
\makebox[21.6mm][c]{wFBP (\textsc{a})}%
\makebox[21.6mm][c]{IR-TV (\textsc{b})}%
\makebox[21.6mm][c]{\textbf{Ours} (\textsc{g})}\\
\includegraphics[width=\linewidth,interpolate=false,trim={28 0 2 22.5},clip]{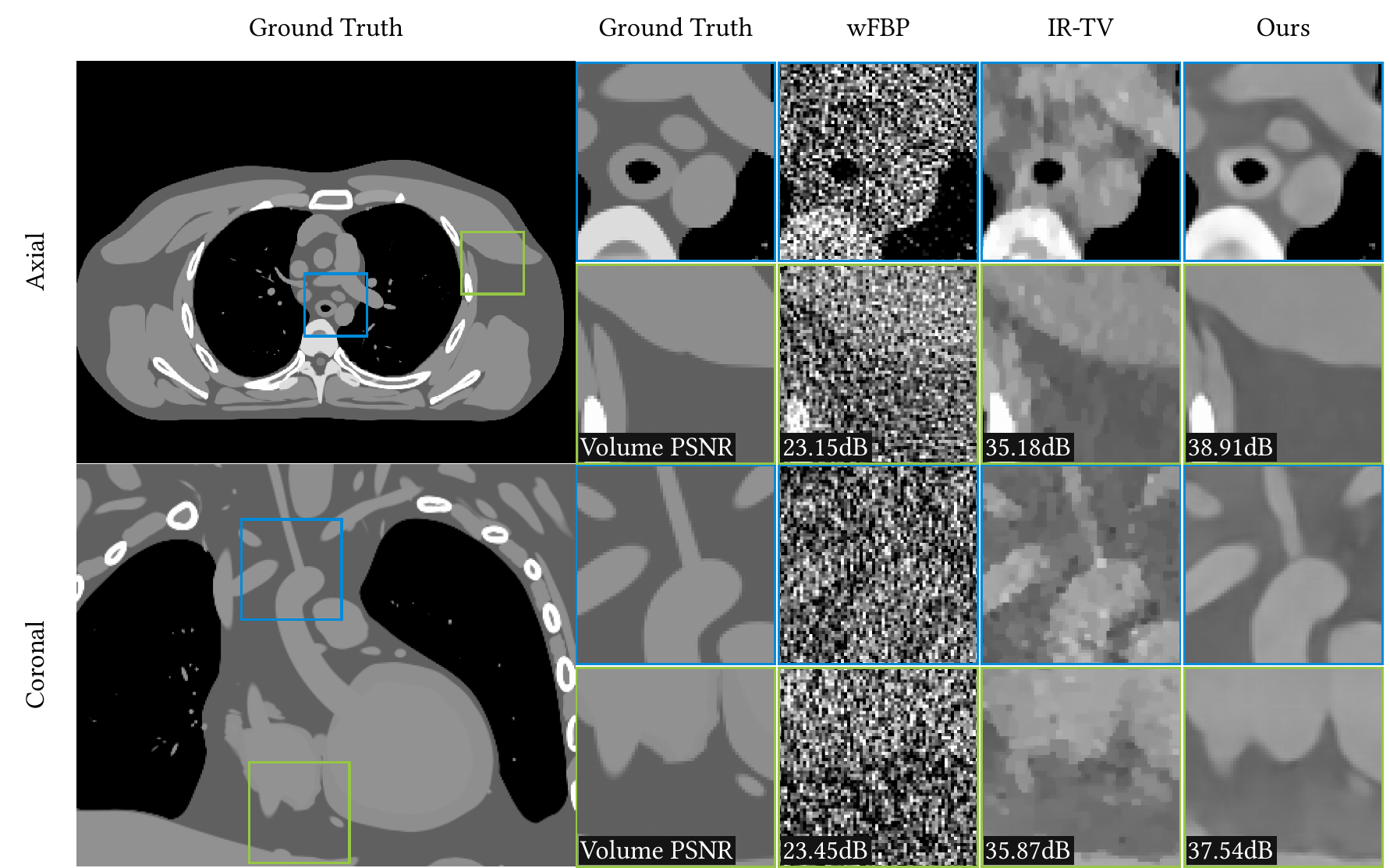}%
\makebox(0,0)[l]{\hspace{-\linewidth}\hspace{1mm}\raisebox{83mm}{\rotatebox{90}{%
  \color{white}\contourlength{0.3mm}\contour{black}{Coronal\hspace{35mm}Axial}%
}}}%
\caption{\label{figXcatLdToLd}%
Low-dose reconstructions of synthetic data using our method trained with low-dose targets.
Display window is set to \mbox{[\tminus400, 400] HU}.
The wFBP and IR-TV results in \autoref{figXcatMainResultsFigure} are duplicated here to facilitate visual comparison.
}
\vspace*{-1mm}
\end{figure*}
}

\begin{document}
\title{Simulator-Based Self-Supervision for\\Learned 3D Tomography Reconstruction}
\author{%
  Onni Kosomaa\thanks{Part of work done during an internship with NVIDIA Research.} \\ Aalto University \\
  \ifemail\texttt{onni.kosomaa@aalto.fi}\else\fi
  \And
  Samuli Laine \\ NVIDIA \\
  \ifemail\texttt{slaine@nvidia.com}\else\fi
  \And
  Tero Karras \\ NVIDIA \\
  \ifemail\texttt{tkarras@nvidia.com}\else\fi
  \AND
  Miika Aittala \\ NVIDIA \\
  \ifemail\texttt{maittala@nvidia.com}\else\fi
  \And
  Jaakko Lehtinen \\ Aalto University and NVIDIA \\
  \ifemail\texttt{jaakko.lehtinen@aalto.fi}\else\fi
}
\maketitle

\begin{abstract}
We propose a deep learning method for 3D volumetric reconstruction in low-dose helical cone-beam computed tomography.
Prior machine learning approaches require reference reconstructions computed by another algorithm for training. In contrast, we train our model in a fully self-supervised manner using only noisy 2D X-ray data. This is enabled by incorporating a fast differentiable CT simulator in the training loop. As we do not rely on reference reconstructions, the fidelity of our results is not limited by their potential shortcomings.
We evaluate our method on real helical cone-beam projections and simulated phantoms.
\JL{Our results show significantly higher visual fidelity and better PSNR over techniques that rely on existing reconstructions.}
When applied to full-dose data, our method produces high-quality results orders of magnitude faster than iterative techniques.
\end{abstract}

\section{Introduction}
\label{sec:introduction}
Computed tomography (CT)
is a versatile medical imaging technique for producing tomographic images of body tissues
  from two-dimensional X-ray projections.
Modern systems reconstruct a 3D volume instead of individual 2D slices,
  so that various cross-sections can be examined easily.
For medical CT scans, the most popular mode of acquisition is moving a point-like radiation source and a 2D X-ray detector along a helical trajectory.
Reconstructing tomographic volumes from such helical cone-beam (CB) data is a challenging inverse problem whose difficulty is further exacerbated by the increased noise inherent to scans taken using low radiation doses.
Yet, as CT uses ionizing radiation, minimizing the dose is of paramount importance when operating with living subjects.

There has been increasing interest in applying machine learning methods to the reconstruction problem.
\JL{Yet, previous approaches all share a potentially significant shortcoming:}
They rely on the results of traditional reconstruction algorithms as training targets (``process the input so that it matches this reference reconstruction''), and often also as model inputs (``improve this existing reconstruction'').
Clearly, if the targets contain systematic errors, the model will inherit those errors.

We present a deep learning method for CBCT reconstruction with an unprecedented combination of desirable properties:
It has the same benefits as other machine learning methods, including high inference speed and the ability to learn from data, but at the same time, it avoids the need for reference reconstructions through the use of a novel simulator-based self-supervised training scheme.

\TK{Similar to prior work,} we base our method on the classical weighted filtered backprojection algorithm (wFBP)~\cite{Stierstorfer04} \TK{and} introduce learned components in crucial points to enable the use of data-driven priors \TK{for reducing} noise and \TK{correcting} any remaining approximation errors.
\TK{Our model reconstructs a full 3D volume in a single forward pass based on the thousands of raw 2D X-rays (i.e., the sinogram) taken from it, guaranteeing tomographic consistency along all axes and improving the quality of coronal and sagittal cross-sections.
As no iterative refinement is required, the technique is very fast.}

\TK{In a crucial step beyond prior work,} we train our model to maximize data consistency in the 2D projection domain. This is enabled by incorporating a differentiable CT simulator in the training loop: 
Intuitively, the output volume is good when simulated X-rays from it look similar to real held-out X-rays from the respective training data scan.
\TK{The training process is self-supervised in the sense that it requires no reference data besides the noisy 2D projections. This means} that the achievable output quality is not limited by the quality of, e.g., clean reference data or a reference reconstruction method.
While projection consistency is often employed by iterative methods (e.g.,~\cite{Sidky08,Kim17}), it has, to our knowledge, not been utilized as a training objective in learned end-to-end reconstruction before. \JL{This may be due to the  formidable memory and computation requirements posed by the training-time backpropagation through the processing of thousands of input X-rays} \TK{in each iteration.} \JL{Indeed, one of our contributions is to show that this is possible in the first place through} \TK{gradient checkpointing, sparse backpropagation, custom CUDA kernels, etc.}

We present a suite of results that demonstrates clear benefit from the self-supervised training of the full sinogram-to-volume model, as opposed to using classical reconstructions as training targets and/or inputs. The benefits are further corroborated by \JL{considerable PSNR improvements} on synthetic phantoms where a ground truth volume is available.
Although our main focus is on low-dose inputs, our method can be applied to full-dose data as well.
In these cases, our method produces high-quality solutions orders of magnitude faster than iterative techniques.

\ifacks %
Project page with supplemental results is available at {\footnotesize\textbf{\url{https://users.aalto.fi/~kosomao1/self-sup-ct/}}}
\fi

\section{Previous work}
\label{sec:previous}

\SL{Mathematically exact tomography reconstruction algorithms for helical CBCT are known only in the limit of infinitely fine discretization and perfect sampling \cite{katsevich2002}.
In the practical case, all known feedforward methods are approximate.
}
This has given rise to a large family of iterative techniques that directly optimize the volume to minimize a projection consistency loss\,---\,difference between synthetic X-rays computed from the volume and the input X-rays\,---\,\TK{employing sophisticated regularization} to deal with the severe ill-conditioning \cite{Sidky08,Kim17}. Unfortunately, iterative techniques are orders of magnitude slower than feedforward methods.

\vparagraph{FBP and wFBP.}
Most non-iterative methods in wide use are based on filtered backprojection (FBP)~\cite{Wang18} and particularly its weighted variant (wFBP)~\cite{Stierstorfer04}. The basic idea is to ``pull'' the input \mbox{X-ray} projections back from the sensor onto the voxel grid along straight lines\,---\,the origin of the term ``backprojection''\,---\,and average the results, after first applying a linear ramp filter to counteract the overrepresentation of low frequencies that the average would otherwise exhibit.
\TK{To handle the challenging CBCT case, methods in this family rebin and reweight the projections to approximate a more tractable projection geometry.}
Despite the much greater efficiency compared to iterative techniques, these approximations cause characteristic artifacts. We seek a feedforward method as efficient as wFBP but without its flaws, and therefore concentrate on prior feedforward methods.

\vparagraph{Machine learning methods.}
A growing body of research applies machine learning to CT reconstruction. To our knowledge, all prior work on feedforward reconstruction is based on supervising a model with the result of an existing (non-learned) technique, typically wFBP.
\TK{Many techniques take an existing low-quality volume and attempt to remove artifacts and noise from it by a neural network that is trained to match a high-quality reference volume \TK{(e.g., \cite{Chen17,Zamyatin22})}. The input and target volumes are typically obtained by running wFBP on matching low-dose and full-dose scans, respectively.}
Another branch of techniques \TK{operates} directly on the raw input projections, but still \TK{supervises} the reconstruction by a wFBP reference (e.g., \cite{Wurfl18,He20}). Some previous works employ the Noise2Noise principle \cite{Noise2noise} in an attempt to mitigate the lack of ideal, noise-free target volumes \cite{Hendriksen20,Wu21,Jing22}. This entails splitting the input projections into two sets and computing the wFBP target from projections that are not used \TK{as} model inputs.
Some techniques employ adversarial losses to mitigate the blurring caused by minimizing MSE \cite{Yang18,Wolterink17,Cohen2018}. %
\TK{The apparent increase in detail comes at the} risk of introducing spurious features not actually present in the input data. All these techniques share the common shortcoming that the systematic errors in the wFBP training targets limit the fidelity of the results.

\vparagraph{Simulator-based supervision.}
Using explicit simulation of the physical process that produces the measurements, many classical inverse problem solvers\,---\,including iterative CT reconstruction methods\,---\,seek a solution that gives rise to the same measurements under the simulation as those observed in reality.
We note that this concept has also been applied in other branches of deep learning. For example, neural radiance fields (NeRF) \cite{Mildenhall2020} make use of differentiable volume tracing operations to seek a view-dependent 3D volume that explains the observed photographs.
Similar ideas have been applied to, e.g., material appearance and illumination recovery \cite{NimierDavid2021,Li2022,Deschaintre2018}, neural control of physical simulations \cite{Hu2019,Liang2019}, neural image restoration through differentiable image corruptions \cite{Chen2018}, and joint design of optics and neural image reconstruction \cite{Tseng2021}.

\section{Our reconstruction pipeline}
\label{sec:pipeline}

\figPipelineOverview

We combine the efficiency of wFBP with the high fidelity of iterative techniques through a careful combination of learned and fixed-function components, trained with a differentiable CT simulator to encourage synthetic projections from the reconstructed volume to match the actual input projections.
Coupled with simple augmentations and a randomized \TK{leave-out strategy}, this \TK{gives rise to} consistent, high-quality reconstructions without relying explicit volume priors.
We will first walk through \TK{our} pipeline in detail, and then describe our self-supervised training setup in \autoref{sec:training}.

The overall structure of our reconstruction pipeline, illustrated in \autoref{figPipelineOverview}, is shared with wFBP (cf. \autoref{sec:previous}). There are four major deviations: a 2D pre-processing neural network for the raw 2D X-ray images, a learned linear ramp filter, \TK{a} differentiable CB backprojection \TK{operator}, and a 3D neural network to \TK{produce} the final volume.
A key benefit of our design is that it can be easily adapted to any acquisition setup with variable number of projections, spacing of the helical trajectory, and radiation doses, with the learned components complementing the fixed backprojection in a data-driven manner.

\subsection{Pipeline walkthrough}

\figNetworkInputOutput

\vparagraph{Geometric setup.}
We target the standard CBCT setting, illustrated in \autoref{figNetworkInputOutput}a, where a radiation point source and a sensor grid travel along a helical trajectory. They capture a sequence of 2D projections that measure the attenuation along each ray within the cone as it penetrates the volume.
\TK{In our experiments, each scan consists of 9,000--15,000 projections stored at 736$\times$64 resolution, } %
\JL{so that each volume reconstruction is based on roughly half a gigapixel of X-ray images.}

\vparagraph{2D network.}
Given a set of projections as an input, we first feed each of them through a learned 2D neural network that outputs a single-channel feature map in the same spatial resolution as the input.
While the network has no other task than to prepare the projections for subsequent processing, we observe that it learns to perform 2D denoising to the inputs (\autoref{figNetworkInputOutput}c).
We visualize the resulting feature map in false color, as it is not guaranteed to be in interpretable units.
While we could output a higher number of feature maps from the model, \TK{we have observed} no benefit \TK{in} doing so.

\vparagraph{Learned ramp filter.}
Next, to prepare for the backprojection operator that transfers the projection information from 2D to 3D, we filter the projections along the cone-beam rows using a learned ramp filter.
This is implemented as a convolution with a one-dimensional kernel that is twice as wide as the input projections.
The filter taps are initialized to the inverse Fourier transform of the desired ramp frequency response, \JL{following wFBP,} but they are treated as learnable parameters during training.
This allows the pipeline to adjust the frequency spectrum of the projections to facilitate the later operations; in practice, we have observed that the ramp filter changes very little during training.

\vparagraph{Differentiable backprojection.}
To transfer the 2D feature maps into the 3D volume, we pass each of them to a fixed-function differentiable backprojection operator
  that accumulates log-space attenuation to all voxels intersected by the cone-beam in question. 
In contrast to wFBP, we perform backprojection directly using cone-beam geometry\,---\,i.e., along lines that connect sensor pixels with the radiation source\,---\,without rebinning to parallel beam projections first.

The backprojected value for a voxel is obtained by first projecting its center onto the 2D sensor using the radiation source as the center of projection, and then interpolating along the sensor's pixel grid.
The projection lines converge onto the radiation source, which causes the local frequency content of the backprojected signal to vary significantly: Close to the source, the projection lines are packed densely, while near the sensor their spacing is sparser. As using a voxel grid fine enough to capture the densest beam bundles is impractical, we carefully anti-alias the result to ensure the backprojected signal can be represented by the voxel grid faithfully, following the standard Shannon--Nyquist sampling and reconstruction theory~\cite{Shannon1949}. In practice, we efficiently approximate the required pre-filtering through \emph{mipmapping}~\cite{mipmap}, a technique common in computer graphics.
\SL{Further details are given in \refappMethod{}.
Following Stierstorfer et~al.~\cite{Stierstorfer04}, we apply a cosine-tapered weight function to detector rows
  and normalize each voxel by dividing the accumulated value by the total weight of contributing backprojections.
Compared to wFBP, our wider taper ($Q=0.8$) and lack of rebinning results in stronger spiral artifacts, but we find them to be easily corrected by subsequent processing.
}

\vparagraph{Learned 3D processing.}
As the final step of the reconstruction pipeline, we pass the volume through a learned 3D network that outputs a voxel grid in log-attenuation space (\autoref{figNetworkInputOutput}c).
As its receptive field is relatively large, it can correct for blur, spiraling, and other artifacts caused by the earlier stages.
\TK{In practice, the voxel resolution of our final reconstruction varies between 576$\times$576$\times$272 and 1024$\times$1024$\times$644, depending on the case.}

\subsection{Network details}
We use U-Nets~\cite{Ronneberger2015}, i.e., autoencoders with skip connections, for the 2D and 3D networks as they have been shown to perform well on a variety of tasks including denoising and removal of image artifacts~\cite{Mao2016b}.
\TK{Our} 2D network architecture \TK{matches the one used by} Lehtinen~\etal{}~\cite{Noise2noise}. The 3D network is similar, except that the intermediate channel counts have been halved to conserve memory, and 3$\times$3 convolution kernels have been replaced with 3$\times$3$\times$3 kernels to enable volume processing.
Network weights were initialized using He initialization~\cite{He2015}.

\section{Training}
\label{sec:training}

\figTrainingPipeline

We now turn to training the learned 2D neural network, ramp filter weights, and 3D neural network.
We train the pipeline in an end-to-end fashion, meaning that only the fidelity of the final reconstruction provides the signal that guides the components to a joint optimum.
\SL{We describe the overall architecture of the self-supervised loss function and training loop in \autoref{sec:selfsup}, deal with photon noise in the training data in \autoref{sec:n2n}, and} \TK{discuss projection simulation and the remaining challenges} \SL{in \autoref{sec:trainingdetails}.}
The \SL{training} process is illustrated in \autoref{figTrainingPipeline}.

\subsection{Simulator-based self-supervision}
\label{sec:selfsup}

To enable training without known reference 3D volumes, we combine a projection consistency loss, similar to many iterative reconstruction techniques, with a leave-out strategy that resembles cross validation: 
  A volume reconstruction is considered faithful if left-out real X-rays look the same as simulated \mbox{X-rays} computed using the same scanner position. 
A key benefit of this approach is that it requires no reference data \TK{in either 2D or 3D}. %

Each training iteration begins by selecting a random slab of the volume from a scan in the dataset, and identifying the set of X-rays whose backprojections overlap with the slab.
The set is then randomly split into a large set of \emph{input projections} and a small set of \emph{target projections} (\autoref{figNetworkInputOutput}b).
The input projections are fed to our reconstruction pipeline, resulting in a 3D volume.
We then compute, for each target projection, a virtual X-ray using the known positions of the radiation source and sensor using a differentiable X-ray simulator (\autoref{sec:trainingdetails}).
The final loss function is the mean squared error between the simulated projections and left-out target projections.
As all components in the pipeline are differentiable, the gradients of the learnable parameters
are computed using standard backpropagation, and used to train the networks with the Adam~\cite{Adam} optimizer.

\subsection{Noisy target projections}
\label{sec:n2n}

A subtle point not addressed in the discussion above is that as we train with real X-ray data, we do not have noise-free projections at hand, i.e., the target projections contain all forms of noise inherent to X-ray imaging. Fortunately, this noise fulfills the requirements for the \emph{Noise2Noise} principle \cite{Noise2noise} to apply: It is zero-mean and uncorrelated between the inputs and outputs, as noise realizations between different X-ray images are independent. Accordingly, the noise ``averages out'' when the model is trained with the noisy targets and $L_2$ loss, and given enough data, the model converges to the same optimum as though it were trained with clean targets.

The requirement that the noise in the model inputs must be uncorrelated with training targets is also the reason behind the leave-out strategy of not using target projections as model inputs.
If this is not met, the quality of the results deteriorates dramatically, as \TK{we demonstrate} in \refappResults{}.
While the leave-out strategy leaves gaps in the set of input projections during training, we have found the impact of these gaps to be negligible as long as the number of target projections is kept small.
For each training iteration, we use approx.~7000 input projections and~12 target projections.

An important detail to consider is that the noise in the acquired 2D projections is zero-mean in photon intensity, but not in log-attenuation because of the nonlinear transformation.
As such, to use noisy training targets, we must compute the $L_2$ loss in photon intensity space.
This, however, has the severe problem that pixels with high photon counts, i.e., low attenuation, have exponentially larger \TK{effective} weight in the overall loss function than highly attenuated pixels, which is at odds with the practice of viewing the results in log-attenuation space.
Therefore, we scale the photon-intensity $L_2$ loss so that each pixel's contribution to the overall loss is proportional to what it would have been if the loss were computed in log-attenuation space.
The resulting loss function is
\begin{equation}
\label{eq:loss}
\mathcal{L}(\hat{X}, \hat{Y})=\big\Vert w(\hat{X})\odot(\hat{X}-\hat{Y})\big\Vert^2_2\ ,\hspace{4mm}w(\hat{X})=1/\hat{X},
\end{equation}
where \smash{$\hat{X}$} and \smash{$\hat{Y}$} are the simulated projection and the training data projection in photon intensity space, respectively, and \smash{$\odot$} denotes element-wise multiplication.
The weighting function $\smash{w(\hat{X})}$ is proportional to the inverse derivative of the exp-transform.
Importantly, we \TK{treat} the gradients of $\smash{w(\hat{X})}$ \TK{as zero} to prevent training from attempting to just minimize this weight~\cite{Noise2noise}.

We have so far assumed that the same projections are used as both inputs to the reconstruction pipeline as well as training targets. %
However, we may have paired projections with different dosages available at training time.
Such paired data can be obtained by, e.g., acquiring higher-dose projections and adding noise to them that simulates a lower-dose scan.
\TK{In this situation, we can use the higher-dose projections as $\smash{\hat{Y}}$ in \autoref{eq:loss}, while still using the lower-dose projections as input.}

\subsection{Projection simulation, augmentations, and feasible implementation}
\label{sec:trainingdetails}
The projection simulator follows the underlying physical measurement process: For each pixel in the projection, we march a ray through the volume and accumulate the attenuation coefficient. We anti-alias the operations through supersampling and trilinear interpolation, and low-pass filter the target projections to control for ringing. To accelerate the learning of high spatial frequencies, we emphasize them by high-pass filtering the difference images in the loss. Similar to many previous works (e.g., \cite{Zeng2000}), we also found it practically beneficial to use our custom mipmapping-based backprojection operator for computing the gradient of the ray-marching operation.

Our setting allows for lightweight geometric data augmentations by simply transforming the associated metadata of sensor device positioning in relation to the volume. For each training sample, we randomize the $z$-axis rotation and add a sub-voxel position jitter to discourage overfitting to orientations or regular grid patterns. We also randomly scale the overall intensity of the projections.

\SL{The amount of computation involved in producing a single reconstruction makes a naive implementation of end-to-end training infeasible:}
  \JL{For example, the intermediate results required for gradient backpropagation through the thousands of invocations of the 2D network} \SL{are far too large to be stored in memory, 
  and even if an infinite amount of memory were available, training would be unacceptably slow.
To make the training process practical and efficient, we use gradient checkpointing, sparsified backpropagation, and a two-level training loop.}
\TK{For efficiency, we implement our differentiable projection and backprojection operations as custom PyTorch~\cite{pytorch} operators using a combination of custom CUDA code and a modified version of the texture lookup function in Nvdiffrast~\cite{nvdiffrast}.}
\SL{Please see \refappMethod{} for further details on training, including signal processing and augmentations.}

\section{Results}

Instead of attempting an exhaustive comparison to all major previous reconstruction algorithms, our main goal in evaluation is, in a common architectural setting, to understand the differences between major \emph{families} of feedforward techniques that differ by the types of input data and the data the model is supervised on. We conceptualize the design space through two axes:
\begin{enumerate}
    \item \TK{\textbf{Model input:} Does the model take an existing 3D volume as input, or does it operate directly on raw 2D projection data?}
    \item \TK{\textbf{Training target:} Is the model trained to match an existing 3D volume, or does the training employ self-supervision based on raw 2D projection data?}
\end{enumerate}
Out of the four possible combinations, previous work covers two: taking a reconstructed volume as input and using another \TK{reconstruction} as a training target (\voltovol), and taking projections as input and using a reconstructed volume as training target (\projtovol). Our approach \TK{(\projtoproj)} is the first to use raw X-rays as both model inputs and training targets without relying on externally-provided reference reconstructions. Compared to the other three choices, we show that this results in significantly improved quality, both \TK{quantitative and qualitative}.

We evaluate our method on both synthetic and real-world \TK{medical chest scans}.
Synthetic data is included to enable quantitative performance measurement\,---\,this is only possible with a known, noise-free volume.
Real-world data enables us to qualitatively confirm that our method scales up to the complexity of real subjects and scanner setups.
To avoid issues inherent to the domain gap, all methods are trained separately on synthetic and real data.
\TK{We run the training in parallel using~8 NVIDIA A100 GPUs for~2.5 days (480 GPU hours) for the synthetic dataset, and 8 days (1536 GPU hours) for the real dataset to accommodate for the higher resolution.
See \refappEvaluation{} for details on comparison methods, datasets, and metrics.}

To facilitate visual comparison, we slightly blur the 3D reconstructions produced by our method to better match the visual look of the ground-truth and wFBP results using a hand-tuned Lanczos filter.
Full versions of result images, including neighboring slices and an interactive HTML viewer, are available as supplemental material. 
In addition to the results presented below, \refappResults{} contains further experiments including validation of the self-supervised training setup, target exclusion, and photon-space loss function, as well as additional result images. %

\subsection{Low-dose inputs}
\label{sec:lowdosesynthetic}

\tabXcatMainResultsTable

\TK{We begin by presenting results using low-dose inputs (10\% of full dose) and full-dose targets (where applicable).
\autoref{tabXcatMainResultsTable} reports the PSNR and RMSE for a representative subset of the methods, computed against noise-free synthetic ground truth volumes, and \autoref{figLdctMainResultsFigure} shows example reconstructions on real data (omitted methods behave consistently with the PSNR measurements). See \refappResults{} for the corresponding reconstructions using synthetic data.}

\vparagraph{Baselines.}
We establish baselines with two traditional methods, wFBP (\configA) and total variation regularized iterative reconstruction (\mbox{IR-TV} \cite{Sidky08}, \configB), \TK{as well as} the machine learning technique RED-CNN (\cite{Chen17}, \configC). Neither wFBP nor \mbox{IR-TV} make use of training data, except that \TK{we tune} the parameters of \mbox{IR-TV} to maximize PSNR over the synthetic dataset. As seen in \autoref{figLdctMainResultsFigure}, wFBP struggles with low-dose inputs, particularly in the soft tissue windows. While \mbox{IR-TV} achieves a significantly higher PSNR and a noticeably better visual result on real data, both techniques' reconstruction quality is limited in the low-dose regime.

RED-CNN is the simplest point in our design space for learned methods: It processes 2D slices of low-dose wFBP reconstructions and uses 2D slices of \SL{full-dose} wFBP reconstructions as training targets. Due to the focus on 2D slices, the method cannot exploit 3D structure. Regardless, it improves PSNR by several decibels over the baseline IR-TV. Finally, to control for the effects of the network architecture, we implemented a version of RED-CNN with our 2D \TK{U-Net} structure (\configD), further improving the result by \PSNRdiff{\psnrFreectUnetTwoDSupldhd}{\psnrFreectRedcnnTwoDSupldhd}\,dB.

\figLdctMainResultsFigure

\vparagraph{\Voltovol and \projtovol.}
We now move on to machine learning techniques that process and produce 3D volumes. To take a minimal step from \configD and gauge the potential benefit of providing the model a view to the full volume instead of just slices, we trained a denoiser network based on our 3D U-Net architecture (\configE) using full-dose wFBP reconstructions as  targets and wFBP low-dose reconstructions as inputs (\voltovol). Effectively, this is a 3D version of RED-CNN\TK{, similar to \cite{Zamyatin22}}.
This improves PSNR further by \PSNRdiff{\psnrFreectUnetThreeDSupldhd}{\psnrFreectUnetTwoDSupldhd}\,dB,
  confirming that access to 3D structure improves denoising results numerically. As seen in \autoref{figLdctMainResultsFigure}, \configE achieves a relatively good visual result on real data, with some blur and noise remaining.
  \refappResults{} highlights the visual importance of denoising over the full 3D volume instead of processing slice-by-slice, which results in strong artifacts along the non-axial cross-sections.

Next, in \configF, we investigate whether additional performance is available by switching to using \TK{2D} projections as model inputs instead of volumes reconstructed by wFBP\TK{, similar to previous end-to-end approaches \cite{Wurfl18,He20}}. This configuration consists of the entire pipeline depicted in \autoref{figPipelineOverview} that is trained using \SL{full-dose} wFBP reconstructions as targets. Interestingly, though the model's inputs are \TK{richer} than those of \configE, \TK{quantitative} performance decreases noticeably.
\JL{This suggests that the beginning of our pipeline is, up to and including backprojection, sufficiently different from wFBP to make the task of matching the full-dose wFBP target volume harder to achieve.}

\vparagraph{Our method: \projtoproj.}
Up to this point, the training targets for all methods have been  full-dose reconstructions made using wFBP. %
We now turn to our full method, \configG, that uses low-dose projections as inputs and is trained using \SL{full-dose} projections as targets as described in \autoref{sec:selfsup}.
As seen in \autoref{tabXcatMainResultsTable}, this improves PSNR significantly to \PSNRprint{\psnrProjnetUnetThreeDSelfsupldhd}\,dB,
  surpassing the best comparison method by \PSNRdiff{\psnrProjnetUnetThreeDSelfsupldhd}{\psnrFreectUnetThreeDSupldhd}\,dB.

Our reconstructions on real data (\autoref{figLdctMainResultsFigure} and supplemental \TK{material}) exhibit clearly the best visual quality of the comparison methods.  As seen in the 2nd column from the right, our main competitor \configE suffers from residual noise, particularly in the difficult soft tissue regions.
\TK{This is because the low-dose projections of our real-world dataset were constructed by adding noise on top of full-dose projections (\refappEvaluation{}), so that there is a common noise component between inputs and outputs that the 3D network learns to partially preserve.}
Our method removes the target projections from the set of input projections in each training step\TK{, avoiding} this issue. See \refappResults{} for a targeted study on correlated noise and target exclusion.

\vparagraph{Validation: \voltoproj.}
Finally, we verify that the dramatic increase is not due to only the self-supervised loss but an emergent property of the combination of the pipeline and training approach. For this, we constructed the novel \configH variant that uses wFBP volumes as inputs and processes them using our 3D U-Net, but trains the 3D \TK{U-Net} using our self-supervised loss in the projection domain (\voltoproj). The result, \PSNRprint{\psnrProjnetUnetThreeDSupldhd}\,dB, is significantly worse than our full approach. This and the previous results indicate that using raw projection inputs and supervising in the projection domain are indeed greatly beneficial when done together, but not in isolation.

\subsection{Full-dose inputs}

\figHdResults

Owing to the projection-domain loss and leave-out strategy, our method can be trained to operate on full-dose scans as well.
\autoref{tabXcatMainResultsTable} shows that our method achieves the best numerical results compared to wFBP and \mbox{IR-TV}. Other comparison methods are not applicable in this situation, as it is not possible to construct training targets from the same projections that are also used for the inputs without significantly decreasing data efficiency.
As shown in \autoref{figHdResults}, our reconstructions are of clearly higher fidelity than those of the baselines.

\vspace{-1mm}  %

\section{Discussion and future work}

\vspace{-1mm}  %

We have shown that self-supervised training can be highly beneficial in helical CBCT reconstruction, and believe that the idea of combining projection simulation with end-to-end machine learning could be applied in a range of other tomography setups and other inverse imaging problems.

There are also several specific improvements that could be made in the CBCT case.
Most importantly, our training-time model of the imaging setup, i.e., generation of simulated projections from reconstructed volume, is fairly simplistic.
For example, we do not utilize tube current information that determines the exposure for each X-ray and has an effect on the magnitude of noise. %
We also do not currently attempt to reproduce effects such as beam hardening (selective attenuation of low-energy photons), scattering, and metal artifacts. %
We believe that using uncorrected raw projection data and simulating these effects during loss computation could lead to further significant improvements,
  as these artifact-inducing effects are presumably easier to simulate than to remove directly.

\ifarxiv\else %
\vparagraph{Broader impact.}
We do not expect negative societal impacts from advances in medical imaging. %
\fi

\ifacks
\vparagraph{Acknowledgments.}
We thank Timo Aila for discussions and comments on the manuscript;
Samuli Siltanen, Alexander Meaney, Antti Korvenoja, Mika Kortesniemi, Juha J\"arvel\"ainen and Jussi Hirvonen for discussions;
David Luebke and Kimmo Kaski for general support;
and
Tero Kuosmanen, Janne Hellsten and Samuel Klenberg for compute infrastructure support.
The project made use of computational resources provided by the Aalto Science-IT project and the Finnish IT Center for Science (CSC).
Onni Kosomaa's NVIDIA Research internship notwithstanding, he was funded by the Finnish Center for Artificial Intelligence (FCAI), a part of the Academy of Finland Flagship programme.
\fi

{\small %
\bibliographystyle{IEEEtran}
\bibliography{paper}
}

\ifappendix
\newpage
\appendix
{\LARGE\bf Appendices}
\newcommand{\refpaperNtoN}[0]{\autoref{sec:n2n}}
\newcommand{\refpaperLoss}[0]{\autoref{eq:loss}}
\newcommand{\refpaperProjSim}[0]{\autoref{sec:trainingdetails}}
\newcommand{\refpaperResultTable}[0]{\autoref{tabXcatMainResultsTable}}
\newcommand{\refpaperLdctMainResultsFigure}[0]{\autoref{figLdctMainResultsFigure}}
\section{Evaluation details}
\label{app:evaluation}

\subsection{Datasets}

\afterpage{\tabDatasetTable}
The specifications of our two datasets are listed in \autoref{tabDatasetTable}.

\vparagraph{Real-world data.}
For the real data experiments we use the Low Dose CT Image and Projection Dataset (LDCT)~\cite{ldct}, built at Mayo Clinic, from which we use 47 chest scans captured on Siemens scanners.
Each scan contains full-dose projections with various corrections (e.g., for beam hardening, scattering, nonuniformity) applied to them by the scanner manufacturer.
In addition, each scan has a corresponding set of simulated low-dose (10\% of full dose) projections created by adding noise on top of the full-dose projections.
It is important to note that the low-dose and full-dose data are correlated, as they both contain the full-dose noise.
Each scan also has a reference 3D reconstruction computed by the scanner manufacturer, but these references are noisy, so we cannot perform numerical comparisons against them.

\vparagraph{Synthetic data.}
We generate a synthetic helical cone-beam dataset using the XCAT CT projection simulator~\cite{Segars08}.
We simulate full-dose and low-dose scans of each phantom.
We chose the scanner parameters to match those of the real-world dataset with the exception that we do not use a flying focal spot.
Because the projections in the real-world dataset have beam hardening correction applied, we simulate monochromatic radiation with an energy level of 80\,\keV{} to avoid beam hardening effects in the synthetic data as well.
Following the real-world setup further, we simulate tube current modulation that attempts to keep the photon Poisson noise roughly consistent throughout the scan.
To produce high-quality ground truth volumes, we export the voxel output from XCAT at 16$\times$ our target resolution and downscale it using a 16$\times$16$\times$16-voxel box filter, yielding a total of 4096 samples per output voxel.

Our loss function (\refpaperLoss{} of the main paper) operates on photon intensities that we obtain directly for the synthetic XCAT projections.
The LDCT data contains only log-attenuation values, and we calculate the corresponding per-pixel photon intensities by inverting the log transform.
The content in the XCAT dataset occupies a smaller area in the axial plane than the LDCT dataset; we use a reconstruction diameter of 452\,mm for XCAT and 600\,mm for LDCT.

\subsection{Metrics}

As reconstruction results are typically viewed in the log-attenuation space, we report root mean squared error (RMSE) and peak signal-to-noise ratio (PSNR) computed from log-attenuation results.
We compute both metrics without clipping or quantization, assuming a display window of~2000 Hounsfield Units (HU).
This window fully covers the variation in XCAT data.
We compute the metrics over the full 3D volume instead of, e.g., averaging over individual 2D slices.
In all cases, evaluation is performed using a separate validation set, i.e., a subset of scans that were not shown to the model during training.

\subsection{Comparison methods}

\newcommand{\lpparam}{c}
We compare our reconstructions with three previous methods: wFBP~\cite{Stierstorfer04}, RED-CNN~\cite{Chen17}, and total variation regularized iterative reconstruction (IR-TV)~\cite{Sidky08}. 
We use FreeCT~\cite{Hoffman16} as the wFBP implementation.
Our RED-CNN re-implementation follows Chen et al.~\cite{Chen17}:
  It has five 5$\times$5 encoder convolution layers followed by five 5$\times$5 decoder convolution layers, each with~96 output channels,
  and the inputs of every other encoder layer are summed to the outputs of the corresponding decoder layers.
The number of trainable parameters is 668k.

For IR-TV, we use a straightforward implementation that employs the same high-quality projection and backprojection operators as our reconstruction pipeline
  and performs the optimization using Adam~\cite{Adam}.
In addition, we weight the projected rays according to their approximate
  noise level using a Poisson noise model~\cite{Thibault07}.
The initial reconstruction is computed using wFBP,
  after which it is optimized iteratively according to a loss function that involves projection consistency and TV regularization terms.
Each iteration uses progressively more simulated projections,
  and the final updates are based on all available projections.
This increases efficiency by making the early iterations
  faster while still being sufficiently accurate.
In total, 210 iterative volume updates are performed during reconstruction.
As is customary, we low-pass filter the input projections to minimize ringing~\cite{Kunze05}.
The low-pass kernel weights $[0.15, 1.0, 0.15]$ and 
  the TV regularization term strength of $2\cdot10^6$ 
  are chosen so that they maximize PSNR on synthetic data.

\subsection{Reconstruction speed}

In \refpaperResultTable{} of the main paper, we report FreeCT~\cite{Hoffman16} runtime for wFBP.
FreeCT is a GPU-accelerated reconstruction library, but our optimizations make our reconstruction pipeline roughly 3$\times$ faster.
All times were measured on a single NVIDIA RTX 3090 GPU with 24\,GB of memory.
Timings for methods that use wFBP-reconstructed volumes as input (e.g., RED-CNN) include the initial wFBP execution.
The IR-TV execution time should be considered a rough estimate, 
  as we did not aim for an optimal speed vs.~quality tradeoff; we simply increased the iteration count until the PSNR leveled off.

\section{Additional results}
\label{app:results}

\subsection{Qualitative results for synthetic data}
\figXcatMainResultsFigure
\figXcatTwoDvsThreeD

\autoref{figXcatMainResultsFigure} shows example reconstructions of synthetic data, corresponding to the ``Low-dose inputs'' column in \refpaperResultTable{} of the main paper.
The synthetic dataset is somewhat less challenging than the real-world dataset, especially in terms of fine details such as the pulmonary alveoli.
Nevertheless, the differences in visual quality between the different methods are similar to the real-world results shown in \refpaperLdctMainResultsFigure{} of the main paper.
Note that \configE performs better on synthetic data than real-world data, because the noise in the low-dose input projections is uncorrelated with the full-dose target projections.

\autoref{figXcatTwoDvsThreeD} further highlights the importance of performing denoising over the full 3D volume.
2D denoisers, such as RED-CNN, construct the 3D volume slice-by-slice, which results in strong artifacts along the non-axial cross-sections.
Our 3D network, on the other hand, produces a consistent 3D volume by virtue of operating across all three axes simultaneously.

\subsection{Using low-dose projections as training targets}
\figXcatLdToLd

Thus far, we have used full-dose projections as training targets for all learning-based methods to maximize their result quality, as we have paired projections with different dosages available in both of our datasets.
In practice, however, it may be desirable to train exclusively using low-dose projections due to the challenges associated with data collection.
\autoref{figXcatLdToLd} shows a preliminary experiment, where we trained our method (\configG) using synthetic low-dose projections as both inputs and targets, without making any other changes to the training setup.
Even though the results are of lower quality than those obtained using full-dose targets (\configG in \autoref{figXcatMainResultsFigure}), they are still superior compared to the two baselines.
Given that low-dose training targets provide less information about the true 3D volume than full-dose projections, we suspect that the reconstruction quality could benefit from increasing the amount of training data in this configuration.

\subsection{Importance of computing the loss in photon space}

To gauge the effect of our photon-space loss (\refpaperLoss{} of the main paper), we trained variants of our method using $L_2$ loss in log-space instead.
The log-transformation is nonlinear and therefore skews the mean of the targets, violating the zero-mean noise requirement of Noise2Noise training \cite{Noise2noise}.
When training with full-dose targets, we did not observe significant differences in the numerical results, suggesting that the skew is small compared to the observed photon counts.
However, when training with low-dose targets, the photon counts are lower, and the log-transformation skew is relatively larger.
With low-dose targets we observed a drop of \PSNRdiff{\psnrProjnetUnetThreeDSelfsupldld}{\psnrProjnetUnetThreeDSelfsupldldLogLoss}\,dB compared to our photon-space loss function.
Visual inspection confirmed that the results overestimated the attenuation, as the loss function was skewed towards higher log-space attenuation values.

\subsection{Correlated noise and target exclusion}

To highlight the importance of not having correlated corruptions in the inputs and targets (\refpaperNtoN{} of the main paper), we performed an experiment where we trained our pipeline without excluding the target projections from the set of input projections.
In this case, the input noise is correlated with the target projections and the networks learn to pass the noise on to the reconstructions.
This leads to significant degradation in reconstruction quality, as illustrated in \autoref{figTargetExclusion}.
Note that the noise present in the real-world reconstructions for the supervised comparison methods (e.g., \configE in \refpaperLdctMainResultsFigure{} of the main paper) also stems from this effect: The full-dose noise is present in both the low-dose and full-dose data, so there is no incentive for the networks to remove it.

\figTargetExclusion

\subsection{Experiments with noise-free inputs and targets}

To validate the correctness of our self-supervised training setup, we performed additional experiments with noise-free synthetic data.
If both input and target projections are noise-free, our method converges to a virtually perfect result when using either the supervised or the self-supervised loss.
This confirms that our networks are able to correct for artifacts resulting from the volumetric backprojection in a data-driven way,
  and that the self-supervised loss achieves results on par with the supervised loss when having access to synthetic ground truth reconstructions.

\section{Model and training details}
\label{app:method}

\subsection{Network architectures}

The architecture and channel counts of our projection and volume networks are shown in \autoref{tabNetworkTable}.
All convolution layers use Leaky ReLU \cite{Maas2013} activation function with $\alpha=0.1$,
  except for the final 1$\times$1 convolution that has linear activation.
Size-preserving padding is used at all convolutions.
In total, there are 988k and 742k trainable parameters in the projection and volume networks, respectively.
We have observed that halving the channel count of the 3D network hurts the numerical results significantly, whereas halving the channel count of the 2D network has a fairly small impact.
Hence, it appears that increasing the channel count of the 3D network could improve the results further.

To avoid dependence on the physical units used in the data, we compute the mean and standard deviation of the projection and volume data,
  and scale and bias the network inputs and outputs so that the networks operate on zero mean and unit variance data on average.
  In other words, we define
  $y = f[(x-\mu_x)/\sigma_x] \cdot \sigma_y + \mu_y$, where
  $f$ is the 2D or 3D network, 
  $x$ and $y$ are the corresponding input and output,
  and $\mu$ and $\sigma$ are the mean and standard deviation of $x$ and $y$, computed over the training set.

\tabNetworkTable

\subsection{Gradient checkpointing}

Normally, when training neural networks, intermediate results are stored in memory in the forward pass so that they can be reused in the backward pass for backpropagation.
For example, a convolution layer needs to remember its input in order to compute gradients for the convolution weights in the backward pass.
\emph{Gradient checkpointing} refers to a technique where the layer inputs are stored only at certain checkpoints.
When an input is needed in the backward pass, the necessary forward operations are re-run from the closest checkpoint to regenerate them.
As such, additional computation is expended but memory usage is lowered.

In our system, we use checkpointing in both the 2D projection network and the 3D volume network. 
The 2D projection network is run for all $\sim$7000 input projections when computing a single reconstruction.
The intermediate data for each invocation consumes approximately 100MB of memory,
  and thus storing all the intermediate results would require 700GB of GPU memory in total.
To avoid this situation, we store no intermediate results at all in the forward passes of the projection network.
Instead, during backpropagation we regenerate the inputs of each layer by running the network once for each input projection in turn.
This increases the cost of training the projection network by $\sim$33\%:
  In addition to the usual forward and backward pass (typically twice as costly as a forward pass), an extra forward pass is required in between.

The volume network is run only once per reconstruction, but it operates on a very large 3D tensor representing the full-resolution volume.
The intermediate results would again overrun GPU memory, which necessitates gradient checkpointing in the highest-resolution layers
  where the intermediate data is at its largest.
\autoref{tabNetworkTable} indicates the checkpointing blocks that we use for the volume network in the $N_\mathrm{3D}$ column.
Layers in the checkpointing blocks need to be re-run once during backpropagation to regenerate the missing inputs,
  and the increase in computation workload remains below 33\%.

\subsection{Sparse backpropagation using a two-level training loop}

A naive training loop would yield only a single weight update step per reconstruction.
Considering the amount of computation required, this seems wasteful.
We implement a two-level training loop that decouples the number of weight updates from the number of reconstructions, at the expense of slightly approximate reconstructions in the forward pass during training.

After choosing a random scan and $z$ range from training dataset in the outer loop,
  we split the $\sim$7000 input projections and 96 target projections into eight buckets.
For each bucket, we compute and store a partial reconstruction volume by running the input projections through the 2D projection network and backprojecting the results into a voxel grid.
The 3D volume network is not run at this stage.
The resulting partial volumes have the same size as the final reconstruction volume, but each of them only contains the contributions of 1/8\textsuperscript{th} of the input projections.

Then, in an inner loop, we run a training step for each of the eight buckets.
We pass the input projections in the chosen bucket through the 2D processing stage and backprojection, sum the result with the previously calculated partial volumes for the seven other buckets, pass the result through the 3D processing stage, calculate the loss, backpropagate the gradients, and update the weights.
The next iteration of the inner loop uses the updated weights of the 2D projection network, ramp filter and 3D volume network.
As such, we can perform~8 weight updates at the cost of running the 2D and backprojection stages twice and the 3D stages~8 times, with a slight increase in memory usage.
Our scheme represents a form of sparse backpropagation, as only a subset of the input projections are allowed to contribute to the backpropagated gradients in each training iteration.

The partial volumes computed before the inner loop get progressively more out of sync with the most up-to-date projection network weights,
  which limits the number of useful iterations in the inner loop\,---\,%
  we found eight iterations to be optimal in our case.
In our experiments, the two-level training loop converged approximately 3$\times$ faster than the naive training loop, measured in wall-clock time.

\subsection{Training details}

We distribute the training to multiple GPUs in a way that requires at least as many GPUs as the batch size. Each reconstruction in a minibatch is assigned to one or more GPUs. The input projections, 2D processing stages, and backprojection are distributed equally among the assigned GPUs in a data-parallel fashion. The 3D processing, on the othe hand, is implemented in a model-parallel fashion by dividing the input and output channels of each 3D convolution into equal-sized chunks and splitting the associated workload among the assigned GPUs.

The network weights were optimized using the Adam optimizer~\cite{Adam} with learning rate $\lambda=10^{-4}$, $\beta_1=0.9$, $\beta_2=0.99$, $\epsilon=10^{-8}$, and batch size 4. An exponential moving average of the network weights with a decay factor of $0.99$ was tracked during training and used for all evaluations. We trained the networks for a total of 40\,000 gradient steps, corresponding to 5000 outer loop iterations in the two-level training loop.

\subsection{Differentiable backprojection}

\figPrefilter

\autoref{figPrefilter} illustrates the need for variable-size prefilter in the projection domain.
As per standard Shannon--Nyquist sampling and reconstruction theory~\cite{Shannon1949}, 
the input signal (here, a 2D X-ray projection) needs to be prefiltered before sampling to remove spatial frequencies too high to be representable by the output signal (here, the voxel grid).
With cone-beam geometry, this is non-trivial because the projection diverges and the sensor pixel pitch relative to voxel size changes with distance to the X-ray source.
The required prefilter bandwidth is determined by this relative pixel pitch, and therefore it changes across the volume as well.
If the prefilter size was kept constant and determined, say, according to the center of the volume, voxels close to the X-ray source would prefilter too little and exhibit aliasing, whereas voxels close to the detector would prefilter too much causing blur.
In theory, every 3D voxel would need to sample the 2D projection using a dedicated prefilter kernel in order to account for the voxel's unique footprint on the 2D projection.

While supporting a per-voxel arbitrarily-sized 2D prefilter is infeasible, an excellent approximation can be achieved through a variation of \emph{mipmapping}~\cite{mipmap}.
In practice, when preparing to backproject a 2D feature map, we compute a number of progressively smoother versions of it using Lanczos-3~\cite{Lanczos1956} prefilters of increasing support (decreasing bandwidth). %
For each voxel, after the desired prefilter bandwidth has been determined, we reconstruct the backprojected sample via trilinear interpolation from the two prefiltered projections whose filter sizes best match the desired bandwidth.
This is a very close approximation to having a prefilter with arbitrary size and position on the 2D projection.
In our current configuration, we use five prefilter bandwidths chosen to cover the range of filter bandwidths required by the backprojection.
In contrast to standard mipmapping, we keep the resolution of the prefiltered feature maps unchanged.

\subsection{Projection simulation and loss function}
The simulated projections are computed by ray marching through the reconstructed voxel grid using trilinear interpolation.
The samples are taken at stratified random positions along the ray to avoid structured sampling artifacts.
To further prevent aliasing, we simulate the projections at 4$\times$ higher resolution and downsample them by 4$\times$4 using a Lanczos-3 filter. %

As is common in iterative reconstruction techniques~\cite{Kunze05}, we low-pass filter the target projections before computing the loss to adjust the amount of ringing in the results.
We use separable filter kernels $[0.2,1.0,0.2]$ and $[0.05,1.0,0.05]$ for synthetic and real-world data, respectively.

We found that training convergence can be accelerated considerably by emphasizing high spatial frequencies in the pixelwise error images between simulated projections and target projections before computing the mean squared loss.
In practice, we apply a classic (non-learned) ramp filter to the difference images, i.e., the weighted difference inside the norm in \refpaperLoss{}.
Reminiscent of the derivation of the ramp filter in wFBP, this focuses the loss evenly on all frequencies in the volume, whereas low frequencies tend to dominate otherwise.
We use this crucial optimization in all of our training runs\,---\,%
  without it, the convergence speed of high-frequency details was unbearably slow.

\subsection{Augmentations}
Three kinds of data augmentation are applied during training to increase the robustness of the model.
First, we choose a random rotation around the $z$ axis for the voxel grid, ensuring that the networks learn no preferential orientation of features in the $xy$ plane.
Second, we add a random sub-voxel offset for the center of the reconstruction volume to break the alignment of the voxel grid in relation to the scanner geometry.
These augmentations are implemented by perturbing the geometry information in training data; they do not involve, e.g., resampling the projection or volume data itself.

Finally, we scale the log-attenuation values in all input and target projections by a random scalar in range $[0.75, 1.25]$ for each individual reconstruction during training.
This is a valid transformation, because the backprojection and projection operations are linear in log-domain, and it prevents the networks from learning specific attenuation coefficients seen in the training data and exploiting that information when reconstructing previously unseen data.

In early tests with small datasets, we found that the augmentations improved results noticeably through better generalization power.
However, our final training datasets appear to be large enough that the augmentations provide no significant benefit in terms of PSNR or visual quality.
We still chose to use them in all our experiments, as we observed no detrimental effects in enabling them.

\subsection{Custom CUDA kernels}

We developed custom CUDA kernels for fast differentiable backprojection and for the ray-marching projection simulation.

In the backprojection operation, 
  determining the 2D sample point on a projection for a single voxel in the volume requires
  trigonometry, analytic geometry, intersection computation, etc.,
  resulting in a sequence of several dozen mathematical operations.
Implementing each of these using PyTorch native operations is incredibly bandwidth-inefficient:
  For each primitive operation, the operands are fetched from GPU memory, a trivial computation (e.g., addition) is performed, and the results are stored back in memory.
Due to the size of the tensors involved, none of the intermediate results stay in GPU caches,
  making the performance entirely throttled by memory bandwidth so that only a tiny fraction of the arithmetic power of the GPU is utilized.
To overcome the issue, we collect the entire coordinate computation into a single CUDA kernel that also performs the band-limited texture lookup.
PyTorch has a just-in-time compiler for fusing simple operations in certain cases,
  but a custom kernel was the only option for including the mipmapped Nvdiffrast \cite{nvdiffrast} texture lookup as well.

For implementing the ray marching in the projection simulation,
  the natural approach is to step each ray in a loop and accumulate the log-attenuation in a local CUDA variable.
This is not possible using native PyTorch operations.
As such, a relatively simple custom kernel offers an easy way to improve performance of this step considerably.

As noted in \refpaperProjSim{} of the main paper, these operations can be used to compute the gradient of one another.
Even though the correspondence is not exact, this does not matter in practice\,---\,experimentally it appears to be sufficient that both operations produce a high-quality, band-limited output.

\fi

\end{document}